\documentclass[aps,pra,10pt,twocolumn,superscriptaddress,showpacs,preprintnumbers,amsmath,amssymb, longbibliography]{revtex4-1}

\usepackage{graphicx}
\usepackage{color}
\usepackage{dcolumn}
\usepackage{times}
\usepackage{bm}
\usepackage{verbatim}
\usepackage{mathbbol}
\usepackage{amsmath}
\usepackage{amssymb}
\usepackage{yfonts}
\usepackage{physics}
\usepackage{mathrsfs}
\usepackage{upgreek}
\usepackage[caption=false]{subfig}
\usepackage{float}
\definecolor{dark_red}{rgb}{0.75,0,0}
\definecolor{dark_purple}{rgb}{0.75,0,0.75}
\definecolor{dark_blue}{rgb}{0,0,0.75}
\definecolor{dark_green}{rgb}{0,0.60,0}
\usepackage[colorlinks,citecolor=blue,linkcolor=dark_green]{hyperref}

\usepackage{array,kantlipsum}
\usepackage[dvipsnames]{xcolor}
\usepackage{tikz}
\usetikzlibrary{fit}
\usetikzlibrary{
  shapes.multipart,
  matrix,
  positioning,
  shapes.callouts,
  shapes.arrows,
  calc}
  
  \definecolor{myyellow}{RGB}{245,177,0}
\definecolor{mysalmon}{RGB}{255,145,73}

\usepackage{smartdiagram}
\usesmartdiagramlibrary{additions}
\usetikzlibrary{babel}
\tikzstyle{bag} = [align=center]

\newcommand{\longdash}{\textrm{\,---\,}}

\begin{document}

\title{Two-photon double ionization with finite pulses:\\ Application of the virtual sequential model to helium}

\newcommand{\UCFPhys}{Department of Physics, University of Central Florida, Orlando, FL 32816, USA.}
\newcommand{\UCFCREOL}{CREOL, University of Central Florida, Orlando, FL 32816, USA.}
\newcommand{\NIST}{Applied and Computational Mathematics Division, National Institute of Standards and Technology, Gaithersburg, MD 20899, USA.}

\author{Siddhartha Chattopadhyay}
\affiliation{\UCFPhys}
\author{Carlos Marante}
\affiliation{\UCFPhys}
\author{Barry I. Schneider}
\affiliation{\NIST}
\author{Luca Argenti}\email{luca.argenti@ucf.edu}
\affiliation{\UCFPhys}
\affiliation{\UCFCREOL}

\date{\today}

\begin{abstract}
As a step toward the full \emph{ab-initio} description of two-photon double ionization processes, we present a finite-pulse version of the virtual-sequential model for polyelectronic atoms. The model relies on the \emph{ab initio} description of the single ionization scattering states of both the neutral and ionized target system. As a proof of principle and a benchmark, the model is applied to the helium atom using the {\tt NewStock} atomic photoionization code. The results of angularly integrated observables, which are in excellent agreement with existing TDSE (time-dependent Schrödinger equation) simulations, show how the model is able to capture the role of electron correlation in the non-sequential regime, and the influence of autoionizing states in the sequential regime, at a comparatively modest computational cost. The model also reproduces the two-particle interference with ultrashort pulses, which is within reach of current experimental technologies. Furthermore, the model shows the modulation of the joint energy distribution in the vicinity of autoionizing states, which can be probed with extreme-ultraviolet pulses of duration much longer than the characteristic lifetime of the resonance. The formalism discussed here applies also to polyelectronic atoms and molecules, thus opening a window on non-sequential and sequential double ionization in these more complex systems.  
\end{abstract}


\maketitle

\section{\label{sec:intro}Introduction}
Our understanding of electron dynamics in atomic and molecular photoionization processes has drastically improved over the last two decades, driven by the rapid development of coherent extreme-ultraviolet (XUV) and x-ray light sources~\cite{Corkum-NPhys-07, Krausz-RMP-09, Pazourek-RMP-15}. Thanks to the increase in intensity of ionizing pulses at x-ray Free-Electron-Laser facilities (XFELs) as well as with table-top setups, the use of XUV-pump XUV-probe schemes to study unperturbed correlated electronic motion in real time is on the horizon~\cite{Duris-NPhoto-20, Saito-Optica-19, Borrego-Varillas2022}. In these studies, double ionization plays a central role both because of its unique sensitivity to electronic correlation and because it enables the detection of correlated electron pairs in coincidence~\cite{Weber-Nature-00, Bergues-NComm-12, Maansson-NPhys-14}. 

Extensive studies of two-photon double ionization (TPDI) in helium over the past two decades have substantially contributed to the understanding of correlated electron dynamics in atomic physics. Different regimes for the TPDI process contain detailed information on electron correlation-driven dynamics. In the sequential regime, a first photon ionizes the system, generating an intermediate ion, which subsequently absorbs a second photon, thus emitting a second electron. This mechanism, therefore, consists of two independent single photoionization events and it is active only if the absorbed photons are able to ionize the initial target and the residual ion. The sequential mechanism can be described, in its qualitative features, by an independent-particle model.
In the non-sequential regime, the two photons are absorbed in too short a sequence for the system to settle down in a well-defined intermediate state. For this reason, this mechanism can lead to the concerted emission of two electrons even if the energy of a single photon is insufficient to ionize the parent ion, provided that the energy of the two photons together exceeds the double-ionization threshold. 

Since the non-sequential mechanism relies on virtual intermediate states that can be achieved only through multiple single-electron excitations from the reference ground-state configuration, electron correlation plays an essential role in it and must be taken into account in the initial, intermediate, and final states.
Indeed, several perturbative approaches based on lowest-order perturbation theory have demonstrated the role of electron correlation in the non-sequential regime~\cite{Horner-PRA-07, Ivanov-PRA-07, Horner-PRA-08, Horner-PRAR-08, Forre-PRL-10, Nikolopoulos-JPB-01}. 
Only direct numerical solutions of the time-dependent Schrödinger equation (TDSE), however, could settle the residual disputes on the quantitative aspects of this process even for a seemingly elementary system like the helium atom~\cite{Feist-PRA-08, Foumouo-NJP-08, Palacios-PRA-08, Nepstad-PRA-10, Guan-PRA-08, Hu-JPB-04, Colgan-PRL-02, Jiang-PRL-15}. 
The quantitative description of TPDI dynamics entailed by pump-probe experiments requires expensive numerical simulations~\cite{Feist-PRA-08, Palacios-PRA-08, Palacios-PRL-09,  Feist-PRL-09, Palacios-JPB-10, Feist-PRL-11}. When the atom is doubly ionized with extreme ultrashort pulses that contain photon energies close to the sequential TPDI threshold, the distinction between sequential and non-sequential regimes breaks down~\cite{Feist-PRL-09}. 
Time-resolved studies further highlighted the rich attosecond dynamics inherent to TPDI process when doubly excited states (DES) are probed with ultrashort pulses~\cite{Pazourek-PRA-11, Bachau-PRA-11, Zhang-PRA-11, Feist-PRL-11,  Yu-PRA-16, Jiang-PRA-17, Djiokap-JOptics-17, Donsa-PRA-19, Jiang-PRL-20}.

A few decades ago the quantum optics community introduced the idea of two-particle interferometry~\cite{Horne-PRL-89}. Due to the presence of two identical particles in the final state, TPDI is an ideal process to bring two-particle interferometry to atomic and molecular physics. Palacios~\emph{et al.} used TDSE simulations to demonstrate quantum interference in the joint energy distribution of the two photoelectrons generated by TPDI pump-probe processes~\cite{Palacios-PRL-09, Palacios-JPB-10}. 
Thanks to improved XFELs capabilities and advancement in coincidence detection of charged fragments~\cite{Dorner-PhysRep-05}, it should soon be possible to experimentally observe this phenomenon in helium as well as in larger polyelectronic systems.

While the numerical solution of the TDSE is a reliable way to obtain quantitative predictions for TPDI process, it is computationally very expensive. It is natural, therefore, to look for alternatives that can provide semi-quantitative data for at least some observables of interest. In fact, the numerical simulations used to set on firm ground the quantitative aspects of this process allowed some authors to recognize that a simple virtual-sequential model (VSM) with only single-ionization intermediate states and no final-state interaction can already capture several qualitative aspects of TPDI even below the sequential threshold~\cite{Horner-PRA-07, Horner-PRAR-08, Forre-PRL-10, Jiang-PRL-15}. 

In this work we explore this direction further by extending to pump-probe setups the VSM using the finite-pulse formalism~\cite{Galan-PRA-16} in combination with \emph{ab initio} one-photon multichannel ionization amplitudes, which can be obtained with a range of atomic and molecular ionization programs. The finite-pulse virtual-sequential model (FPVSM) is general and it is straightforward to extend it to polyelectronic atoms and molecules. Although the model does not account for one-photon double-ionization amplitudes from the intermediate bound and autoionizing states, it does include the contribution of final autoionizing states, finite-pulse effects, and the interference of multiple ionizing pulses. The total TPDI cross section predicted by the model is in excellent agreement with TDSE simulations both below and above the sequential threshold, reproducing the resonant feature associated to the $sp_2^+$ intermediate DES.  We show that the FPVSM can reproduce the characteristic features of the joint photoelectron energy distribution and energy sharing in helium compared to TDSE simulations at a small computational cost. The FPVSM also reproduces the continuous transition between  non-sequential and sequential features of the TPDI process with extreme ultrashort pulses as well as the two-electron quantum interference. Finally, the FPVSM model is used to probe the modulation in the joint energy distribution in the resonant TPDI of helium with XUV-pulses of duration larger than the lifetime of the atom's brightest DES.  

The paper is organized as follows. In Sec.~\ref{sec:Theo}, we introduce the FPVSM starting from time-dependent perturbation theory. By expanding the field-free resolvent in terms of single-ionization channels, the TPDI amplitude is derived for the general case of polyelectronic atoms. In Sec.~\ref{sec:Resu}, we introduce the close-coupling (CC) expansion to compute the bound-continuum transition matrix elements for the neutral and intermediate parent-ion states. The model is applied to the computation of angularly integrated observables, such as the joint energy  distribution and energy sharing. A two-color pump-probe scheme is proposed to detect two-particle interference and modulation of the joint energy spectra in the resonant TPDI process.   In Sec.~\ref{sec:con}, we present our conclusions. The paper is completed by App.~\ref{app:Recoupling}, which details the full derivation of the FPVSM for arbitrary atoms, and by App.~\ref{app:FrequencyIntegral}, which summarizes for the readers' convenience the use of Faddeyeva's function to evaluate time integrals in two-photon transitions. Atomic units ($\hbar = 1$, $m_e=1$, $q_e=-1$) and the Gauss system are used throughout unless stated otherwise.

\section{\label{sec:Theo}Theory}
The time evolution of a system under the influence of an external field is governed by the time-dependent Schr\"odinger equation (TDSE), which, in the interaction representation is given by,
\begin{eqnarray}
i\partial_t |\Psi_I(t)\rangle = H_I'(t) |\Psi_I(t)\rangle,
\end{eqnarray}
where $H_I'(t) = e^{iH_0t} H_I(t) e^{-iH_0t}$, $H_0$ is the field-free Hamiltonian of the system. In general, the interaction Hamiltonian is the product of a suitable field $\vec{F}(t)$ and operator $\vec{O}$, $H_I(t)=\vec{F}(t)\cdot\vec{O}$. The interaction is often formulated in either velocity gauge, $H_I(t)=\alpha \vec{A}(t)\cdot \vec{P}$, or length gauge, $H_I(t)= \vec{E}(t)\cdot \vec{R}$, where $\vec{A}(t)$ and $\vec{E}(t)$ are the external vector potential and electric field, respectively, $\alpha$ is the fine-structure constant, $\vec{P}=-i\sum_{i=1}^{N_e}\vec{\nabla}_i$, and $\vec{R}=\sum_{i=1}^{N_e}\vec{r}_i$~\cite{Joachain-12}. 
For a system initially ($t\to-\infty$) in its ground state $|g\rangle$, $H_0|g\rangle=\omega_g|g\rangle$, the TDSE can be cast in integral form
\begin{eqnarray}
\label{eq:TSDE}
|\Psi_I(t)\rangle = |g\rangle - i \int_{-\infty} ^{t} H_I(t')|\Psi_I(t')\rangle dt',
\end{eqnarray}
which is the basis for a perturbative expansion in powers of the external field, $|\Psi_I(t)\rangle =\sum_{n=0}^{\infty} |\Psi_I^{(n)}(t)\rangle$. The relevant expressions to compute the photoelectron distribution in TPDI is the second-order perturbative term
\begin{eqnarray}
|\Psi_I^{(2)}\rangle &= & -\int_{-\infty}^{\infty} dt_2 \int_{-\infty} ^{t_2} dt_1 H_I(t_2)H_I(t_1)|g\rangle.
\end{eqnarray}
Without loss of generality, we can assume that the external field is a combination of linearly polarized pulses, with amplitude $F_i(t)$ and polarization $\hat{\epsilon}_i$,
\begin{equation}
\vec{F}(t) = \sum_{i=1}^{N_p}F_i(t) \hat{\epsilon}_i.
\end{equation}
where $N_p$ indicates the number of Gaussian pulses in the expansion of the external field.
The two-photon transition amplitude $\mathcal{A}^{(2)}_{f\gets g}$ to a final stationary state $|f\rangle$, $H_0|f\rangle = \omega_f |f\rangle $, then reads
\begin{eqnarray}
\mathcal{A}^{(2)}_{f\gets g} &=& \sum_{ij\mu\nu}\epsilon_j^\nu\epsilon_i^\mu \int_{-\infty}^{\infty} \hspace{-12pt}dt_2 F_j(t_2) e^{i\omega_f t_2}\hspace{-2pt} \int_{-\infty}^{t_2}\hspace{-12pt} dt_1 F_i(t_1) e^{- i \omega_g t_1} \times\nonumber\\
&\times&\langle f |  O_\nu\, e^{-iH_0(t_2-t_1)} \, O_\mu|g\rangle,\label{eq:2P2}
\end{eqnarray}
where we used the spherical tensor notation, $\hat{\epsilon}\cdot\vec{O} = \epsilon^\mu \mathcal{O}_\mu$~\cite{Varshalovich-88}.
Equivalently, in frequency form
\begin{equation}\begin{split}\label{eq:2PT_FrequencyIntegral}
 \mathcal{A}^{(2)}_{f\leftarrow g}&= -i\sum_{ij\mu\nu}\epsilon_{j}^\nu\epsilon_i^\mu
 \int_{-\infty}^{\infty} d\omega \,\,\tilde{F}_j(\omega_{fg}-\omega) \tilde{F}_i(\omega)\times\\
 &\times\langle f|\mathcal{O}_\nu\,G_0^+ (\omega_g +\omega)\,\mathcal{O}_\mu|g\rangle
\end{split}\end{equation}
where $\tilde{F}(\omega)=(2\pi)^{-1/2} \int_{-\infty}^{+\infty} dtF(t) e^{i\omega t}$, $G_0^+ (\omega) = (\omega-H_0+i0^+)^{-1}$.
For double-ionization final states, $|f\rangle = |\Psi^-_{A,\vec{k}_1\sigma_1\vec{k}_2\sigma_2}\rangle$, where $A$ is the state of the residual grand-parent ion, and $\vec{k}_i$ and $\sigma_i$ are the asymptotic momentum and spin projection of electron $i=1,\,2$. In the present formalism we use the velocity gauge within the dipole approximation.

To proceed further, we introduce the three main approximations that underpin the FPVSM.
First, the expansion of the field-free resolvent is restricted to single-ionization channels,
\begin{equation}
 \mathcal{M}^{(2)}_{fg;\,\nu\mu} \simeq \sum_{a\sigma'} \int d^3k^{\prime} \frac{ 
 \langle \Psi^-_{A,\vec{k}_1\sigma_1\vec{k}_2\sigma_2}| \mathcal{O}_{\nu}|\Psi_{a \vec{k}^{\prime}\sigma'}^-\rangle\,\langle\Psi_{a \vec{k}^{\prime}\sigma'}^-|\mathcal{O}_\mu|g\rangle}{E_g + \omega-E_a-{k'}^2/2 + i0^+},\label{eq:tpdia}
\end{equation}
where $a$ identifies the state of the intermediate ion, with energy $E_a$. With this approximation, we neglect the contribution to double ionization from real and virtual excitations involving both bound and double ionization intermediate states.
Equations~\eqref{eq:tpdia} contains intermediate scattering states with incoming boundary conditions~\cite{Newton2014} for mere convenience.
The dipole transition matrix elements from the ground state to the continuum, $\langle\Psi_{a \vec{k}^{\prime}\sigma'}^-|\mathcal{O}_\mu|g\rangle$, can be computed \emph{ab initio}.
Second, we assume that one of the photoelectrons in the final state retains the same asymptotic state as the photoelectron in the intermediate single-ionization state, i.e., the second photon does not affect the first photoelectron and can only ionize the associated parent ion. This assumption implies that the ionic state of the intermediate scattering state is available at any time, which is justified only if the lifetime of the intermediate autoionizing states is shorter than the pulse duration. Furthermore, the direct one-photon double ionization of the localized part of intermediate autoionizing states is also neglected. 
Finally, when evaluating the dipole matrix element between the intermediate single ionization (SI) state to the final double ionization (DI) state, we neglect the radiative transitions in the continuum, which in principle may be followed by an $(e,2e)$ process.

Despite its clear limitations, this model is nevertheless useful in reproducing many of the essential features of TPDI in the non-sequential regime since the energy of the first photon does not have to match the energy of the intermediate parent ion plus the asymptotic kinetic energy of one of the photoelectrons. Indeed, even below the opening of the sequential threshold, the model predicts a finite value for the double ionization amplitude. The VSM in stationary regime was introduced first by McCurdy \emph{et al.}~\cite{Horner-PRA-07} and subsequently rediscovered in a slightly different form~\cite{Forre-PRL-10}. In helium, this approximation correctly reproduces all the essential features of the TPDI amplitude of the atom below the sequential threshold. Above the sequential threshold, of course, the real sequential mechanism dominates, and hence the model is expected to become even more accurate. 

Appendix~\ref{app:Recoupling} presents a detailed derivation of the TPDI amplitude in terms of one-photon transition matrix elements between bound and single-ionization scattering states of either the neutral or the ionized target atom with well-defined spin and angular-momentum quantum numbers, taking into account electron's exchange symmetry. Here, we report the result for the amplitude to a DI state. The transition amplitude to a state in which the two electrons have well defined asymptotic momenta and spin projections, $\mathcal{A}^{(2)}_{A,\vec{k}_2\sigma_2, \vec{k}_1\sigma_1\gets g}$, can be expressed in terms of amplitudes where the photoelectrons emerge as spherical waves instead, $\mathcal{A}^{(2)}_{A,E_2\ell_2m_2\sigma_2,E_1\ell_1m_1\sigma_1\gets g}$,
\begin{eqnarray}
\mathcal{A}^{(2)}_{A,\vec{k}_2\sigma_2, \vec{k}_1\sigma_1\gets g}=&-&
\hspace{-2pt}\sum_{\{\ell_i m_i\}}\hspace{-2pt}
\frac{e^{i(\sigma_{\ell_1}+\sigma_{\ell_2})}}{i^{\ell_1+\ell_2}k_1k_2} Y_{\ell_1 m_1}(\hat{k}_1)Y_{\ell_2 m_2}(\hat{k}_2)\times\nonumber\\
&\times&\mathcal{A}^{(2)}_{A,E_2\ell_2m_2\sigma_2,E_1\ell_1m_1\sigma_1\gets g},
\end{eqnarray}
where $\sigma_{\ell}$ is a Coulomb phase shift and $Y_{\ell m}(\hat{\Omega})$ are spherical harmonics~\cite{Varshalovich-88}. In the FPVSM, the transition amplitudes to scattering states with spherical photoelectron waves have the following expression in terms of reduced bound-continuum transition matrix elements for the neutral and ionized system and of the external-field parameters,
\begin{eqnarray}
&&\mathcal{A}^{(2)}_{A, E_1 \ell_1 m_1 \sigma_1, E_2 \ell_2 m_2 \sigma_2\gets g}
=\frac{1-\mathcal{P}_{12}}{2i\sqrt{3}} 
C_{\frac{1}{2}\sigma_2,\frac{1}{2}\sigma_1}^{S_A-\Sigma_A}\sum_{La}\Pi_{LS_A}^{-1}\times\nonumber\\
&&\qquad\times\sum_{ij}
\int_{-\infty}^{\infty} d\omega \,
\frac{\tilde{F}_j(E_A+E_1+E_2-E_g-\omega) \tilde{F}_i(\omega)}{E_g+\omega-E_a-E_2+i0^+}
\times\nonumber\\
&&\qquad\times\sum_{M M_a\mu\nu}
C_{L_A M_A, \ell_1 m_1}^{L M} 
C_{L_a M_a,\ell_2 m_2}^{1 \mu}
C_{L_a M_a, 1 \nu}^{L M}
\epsilon_{j}^\nu\epsilon_i^\mu\times\nonumber\\
&&\qquad\times\langle \Psi^{^{2S_a+1}L^{\bar{\pi}_a}(-)}_{A \ell_1 E_1}\|\mathcal{O}_1\|\Phi_{a}\rangle\,
\langle\Psi_{a\ell_2 E_2}^{{^1P^o}(-)}\|\mathcal{O}_1\|g\rangle\,
\label{eq:2PDIAmp}
\end{eqnarray}
where $\mathcal{P}_{12}$ exchanges all the subsequent indices for photoelectrons 1 and 2, $C_{a\alpha,b\beta}^{c\gamma}$ are Clebsch Gordan coefficients and $\Pi_{a}=\sqrt{2a+1}$.
The state $\Psi_{a\ell_2E_2}^{{^1P^o}(-)}$ represents a single-ionization multi-channel scattering state fulfilling incoming boundary conditions. The channel is identified by the quantum numbers of the only open channel with an outgoing spherical photoelectron component state, namely, the parent-ion label $a$, which corresponds to the ionic wave function $\Phi_a$, the asymptotic angular momentum and energy of the photoelectron, $\ell_2$ and $E_2$, respectively, and the total symmetry and multiplicity of the system (${^1P^o}$). The state function $\Psi_{A\ell_1E_1}^{{^{2S_a+1}P^{\bar{\pi}_a}}(-)}$ similarly identifies the scattering state of the ionic system with parity opposite to the intermediate ion's ($\bar{\pi}_a$). 
It is worth pointing out two main features of this expression. First, for Gaussian pulses, the frequency integral can be expressed analytically in terms of the Faddeyeva function~\cite{Faddeeva-1961, Zaghloul-ACM-17, NIST:DLMF}, which can be evaluated numerically at a negligible computational cost~\cite{Galan-PRA-16}. See App.~\ref{app:FrequencyIntegral} for details. Second, thanks to the exchange term, the model can reproduce the interference characteristic of the photoemission of two identical particles~\cite{Palacios-PRL-09}. The fully-differential photoelectron distribution is
\begin{equation}
\frac{dP_A}{d^3k_1d^3k_2}=\sum_{M_A\Sigma_A \sigma_1\sigma_2}\left| \mathcal{A}^{(2)}_{A,\vec{k}_2\sigma_2,\vec{k}_1\sigma_1\gets g}\right|^2.
\end{equation}
After analytically integrating over the solid angles, $\hat{k}_1$ and  $\hat{k}_2$, the joint energy distribution reads
\begin{equation}\label{eq:JED}
\frac{dP_A}{dE_1dE_2}=\sum_{M_A\Sigma_A} \sum_{\{l_im_i\sigma_i\}}\left| \mathcal{A}^{(2)}_{A,E_2{l}_2 m_{2}\sigma_2,E_1 {l}_1 m_{1}\sigma_1\gets g}\right|^2.
\end{equation}

\section{\label{sec:Resu}Two-photon double ionization of Helium}

\begin{figure}[h!]
\includegraphics[width=\columnwidth]{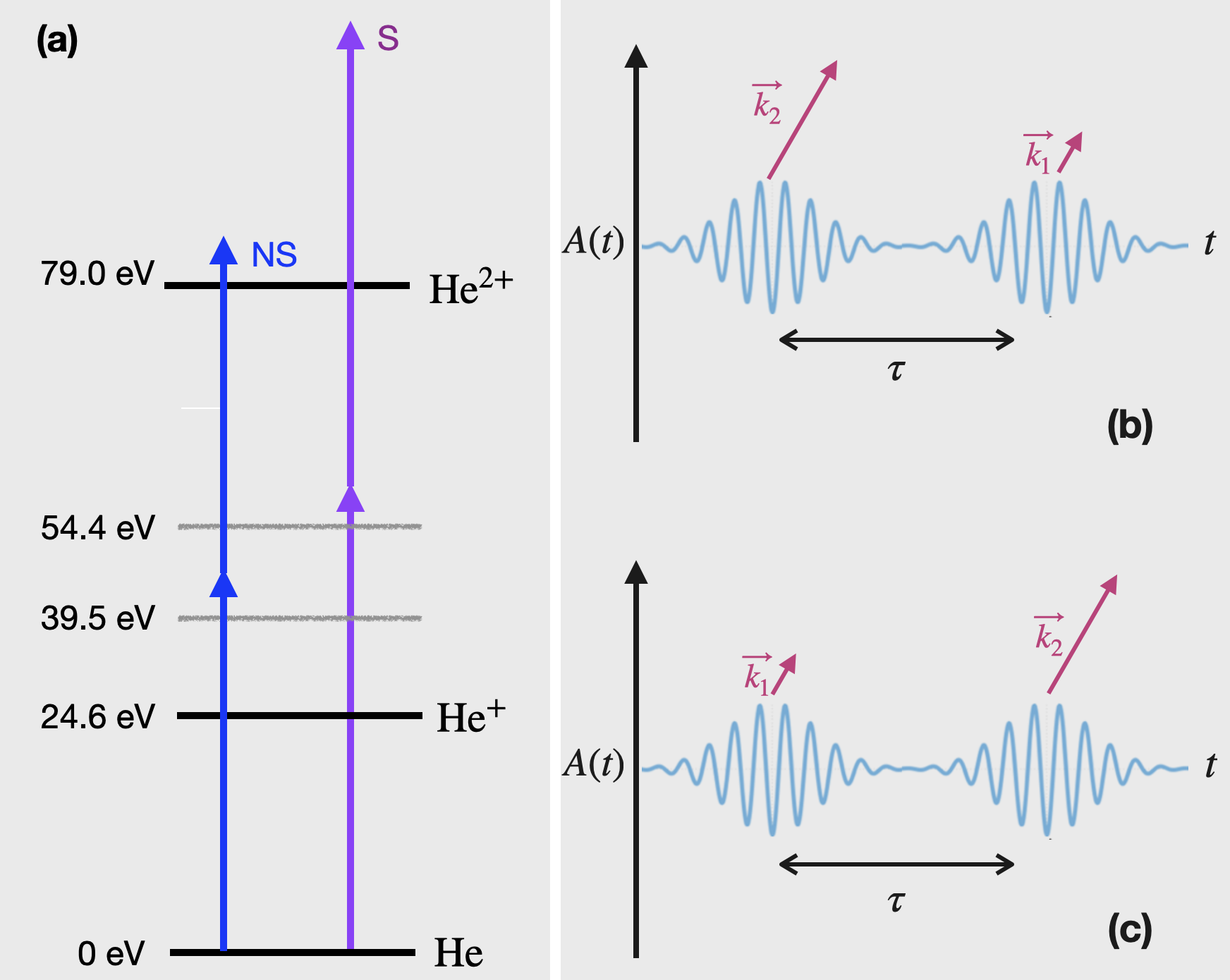}
\caption{\label{Fig:HeEnScheme} XUV-pump XUV-probe scheme for Helium. (a) Non-sequential (NS) and sequential (S) two-photon double ionization. For two identical XUV photons with energy 39.5~eV~$<\omega<$~54.4~eV, double ionization occurs only through the non-sequential mechanism. At energies well above the sequential threshold, $\omega>$~54.4~eV, on the other hand, the sequential mechanism dominates. For finite pulses, two different ionization paths (b,c) lead to a same final state, thus giving rise to characteristic interference fringes in the photoelectron coincidence spectrum.}
\end{figure}

Figure~\ref{Fig:HeEnScheme} illustrates the energy scheme for the TPDI of $\mathrm{He}$ by a sequence of two XUV pulses with relative delay $\tau$.  In the non-sequential and sequential process (online: blue and purple, respectively), a first XUV photon excites the neutral helium atom to the single-ionization continuum. The absorption of another XUV photon ejects the residual electron from the ionic component of the system. The non-sequential process is only possible through the participation of the virtual intermediate states. For the description of TPDI in helium, we have considered the following ionization channels
\begin{eqnarray}
    \mathrm{He}(1s^2) & \xrightarrow{h\nu} & \mathrm{He}^{+}(1s) + e^{-}_{\varepsilon{p}} \xrightarrow{h\nu} \mathrm{He}^{2+} + e^{-}_{\varepsilon{p}}+e^-_{\varepsilon'p} \nonumber \\ 
    \mathrm{He}(1s^2) & \xrightarrow{h\nu} & \mathrm{He}^{+}(2s) + e^{-}_{\varepsilon{p}} \xrightarrow{h\nu} \mathrm{He}^{2+} + e^{-}_{\varepsilon{p}}+e^-_{\varepsilon'p} \nonumber \\ 
    \mathrm{He}(1s^2) & \xrightarrow{h\nu} & \mathrm{He}^{+}(2p) + e^{-}_{\varepsilon\,{s/d}} \xrightarrow{h\nu} \mathrm{He}^{2+} + e^{-}_{\varepsilon\,{s/d}} + e^{-}_{\varepsilon'\,{s/d}} \nonumber.
\end{eqnarray}

The single-ionization amplitudes of helium are computed using the {\tt NewStock} atomic ionization code~\cite{Carette-PRA-13}. We describe the ionization channels through a CC expansion containing the three ionic states: $\mathrm{He}^{+}(1s)$, $\mathrm{He}^{+}(2s)$ and $\mathrm{He}^{+}(2p)$ coupled to an additional electron with an associated \textit{s}, \textit{p} or \textit{d} wave to construct augmented channels with total symmetry $^1S^e$ and $^1P^o$. In addition, to improve the short-range description of the electronic correlation, we included a set of localized two-electron wave functions computed with a Multi-Configuration Hartree-Fock calculation (MCHF) performed using the ATSP2K package~\cite{Fischer-CPC-07}. Both the localized and the continuum channels are expanded in terms of the B-spline basis \cite{Bachau-RPP-01}. We use a simulation box of 300 a.u. \!with $\Delta_r = 0.4$ a.u. \!node separation to build the B-splines functions of degree 7. With the basis depicted above we obtained a helium ground-state energy $E_{gs}=-2.8846$ a.u., which differs by $\approx 0.02$ a.u. from the NIST energy~\cite{NIST_ASD}. In the figures shown in the rest of the paper, we will shift the photon energies so that the position of the ionization thresholds with respect to the ground state coincide with the experimental ones.

\begin{figure}[htb]
\includegraphics[width=\columnwidth]{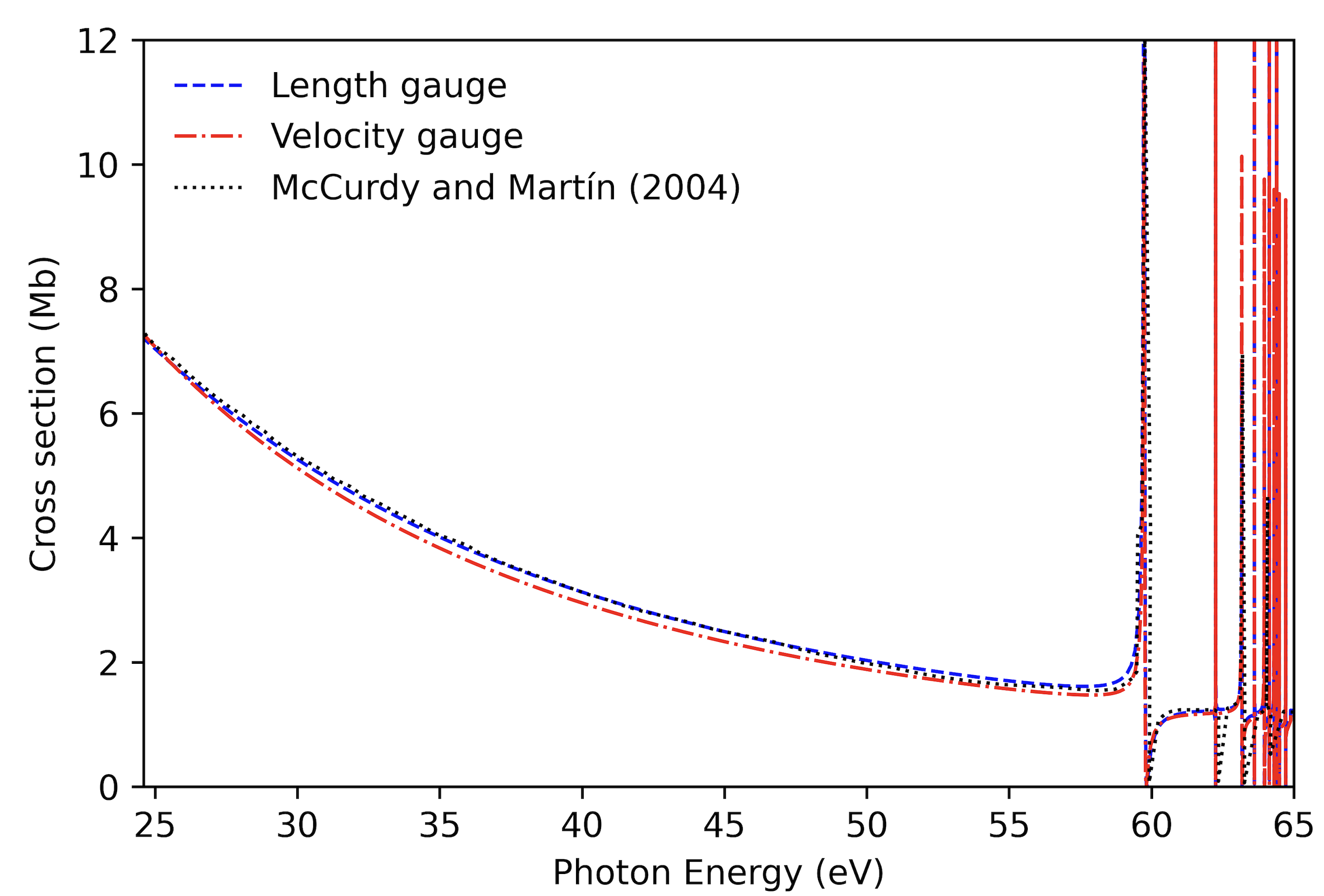}
\caption{
One-photon total photoionization cross-section from the helium ground state computed with {\tt NewStock} in length (blue dashed line) and velocity gauge (red dashed-dotted line). Results from McCurdy and Martín~\cite{McCurdy-JPB-04} (black dotted line) are obtained using the exterior-complex-scaling technique.
}
\label{Fig:HeCrossSection}
\end{figure}

The scattering states corresponding to the single-ionization channels, $|\Psi_{\alpha\vec{k}^{\prime} }^{-}\rangle$ are obtained by solving the Lippmann-Schwinger equation  with incoming boundary condition~\cite{Newton2014}. To test the quality of the intermediate scattering channels defined for the TPDI study, we compare the one-photon total photoionization cross-section with an independent ECS calculation~\cite{McCurdy-JPB-04}. Fig.~\ref{Fig:HeCrossSection} compares our results in the length gauge with the benchmark and provides the {\tt NewStock} velocity gauge as well.  The agreement is excellent, with only a few meV difference in the autoionizing state's position.
The photoionization amplitudes for $\mathrm{He}^+$ are computed with a dedicated numerical one-electron code, which also uses B-splines to represent the radial part of the bound and continuum wavefunctions. We use a B-spline basis defined in a uniform grid with degree 7 and node spacing of 0.4 a.u. with a box size of 300 a.u. With these choices, the one-electron code is able to reproduce the analytical results for the bound-bound and bound-continuum dipolar transition amplitudes to machine precision.

\begin{figure}[hbtp!]
\includegraphics[width=\columnwidth]{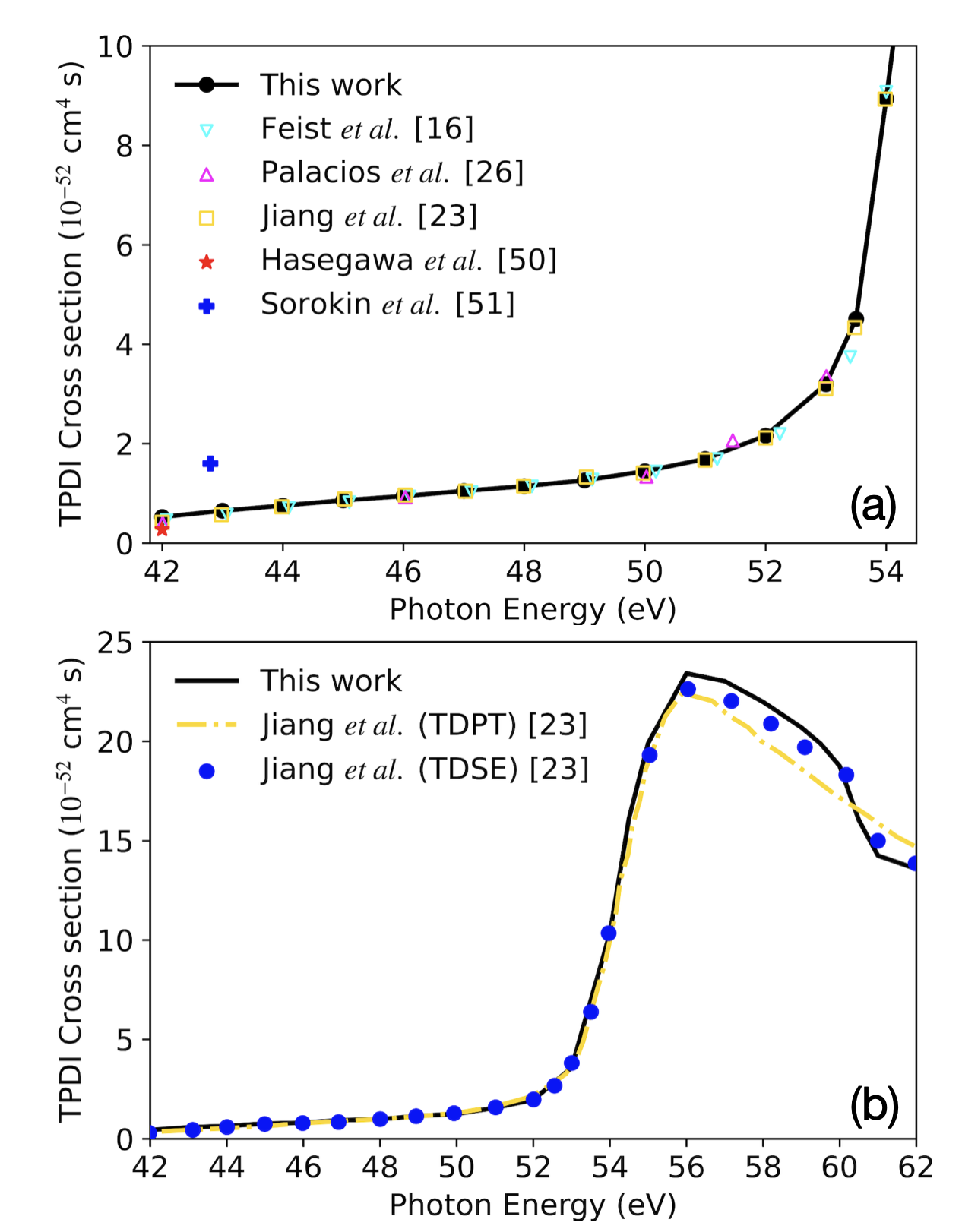}
\caption{\label{Fig:HeTPDI_PICS}
Two-photon double-ionization total cross-section from the helium ground state. (a) Our model results (black circles) with a pulse of  duration 20~fs and intensity $4\times10^{12}$~W/cm$^2$~are compared with previous works. (b) Model results with a pulse duration 4~fs and intensity $4\times 10^{12}$~W/cm$^2$ compared with the results of TDSE and TDPT models of Ref.~\cite{Jiang-PRL-15}.  See text for details.}
\end{figure}
Figure~\ref{Fig:HeTPDI_PICS}~(a) compares several accurate predictions of the TPDI cross section of helium below the sequential threshold, computed by numerically integrating the TDSE using finite-duration pulses with the analytical predictions of the present FPVSM and with the analytical model from~\cite{Jiang-PRL-15}. To compute the TPDI cross-section, we use Eq.~(13) from Ref.~\cite{Feist-PRA-08}, which estimates the TPDI by integrating the total probability over the two photoelectron energies and dividing the result by a form factor proportional to the time integral of the fourth power of the external field. Panel (a) shows our model results with a pulse duration of ~20 fs and intensity of $4\cdot10^{10}$~W/cm$^2$. Along with the theoretical results~\cite{Feist-PRA-08, Palacios-JPB-10,Jiang-PRL-15} previous experimental results~\cite{Hasegawa-PRA-05, Sorokin-PRA-07} are also shown, which can only confirm the cross section order of magnitude far from the sequential threshold. The cross-section in Ref. ~\cite{Feist-PRA-08} (down triangles, cyan online) is obtained by solving TDSE with a pulse of duration 4~fs and intensity $10^{12}$~W/cm$^2$. Ref.~\cite{Palacios-JPB-10} (up triangle, magenta online) reports the total TPDI cross-section computed by solving TDSE with a pulse duration of 3~fs. The TDSE calculation from Ref.~\cite{Jiang-PRL-15} reported in Panel (a) (box, yellow online) corresponds to a pulse duration of 11~fs. The agreement of the model with the numerical simulation is impressive and confirms similar findings based on the virtual sequential model in stationary regime. In particular, the model reproduces the rapid increase in the cross section as the photon energy approaches the sequential threshold already evidenced in~\cite{Jiang-PRL-15}. This rapid increase is due to the enhanced role of the virtual intermediate states when they are close to the resonance condition. 

Figure~\ref{Fig:HeTPDI_PICS}~(b) compares the TPDI cross section across the sequential threshold with the original TDSE \emph{ab initio} calculations and the analytical model from~\cite{Jiang-PRL-15} with duration 4~fs and intensity of $4\times10^{12}$~W/cm$^2$ with different central photon energies. Notice that while the TDPT model from~\cite{Jiang-PRL-15} coincides with ours below the sequential threshold, above the sequential threshold, only our model reproduces the resonance profile due to the excitation of the $sp_{2}^{+}$ state, and is in much better agreement with the result from the TDSE simulation. 

In our calculations we use pulses with Gaussian profile, whereas the TDSE simulations in Fig.~\ref{Fig:HeTPDI_PICS} and the analytical model in~\cite{Jiang-PRL-15} use $\sin^2$ envelopes. As long as the width at half maximum of the two field envelope coincide, we do not discern appreciable differences due to the pulse shape. Furthermore, our model can use a linear combination of an arbitrary number of Gaussian pulses, and hence it can reproduce any pulse shape with any desired precision. For example, a linear combination of as few as five Gaussian functions can fit a $\sin^2$ profile with an error within 0.001 of the field peak value. Since a single Gaussian function already gives results in line with those with a $\sin^2$ profile, however, a calculation with such tailored pulse was not necessary.

\subsection{TPDI joint energy distribution with single pulses}
This section presents the predictions of the FPVSM in the non-sequential regime as well as across the sequential threshold. To illustrate the transition from the non-sequential threshold, at $\omega=39.5$~eV, to the sequential threshold, at $\omega=54.4$~eV, we look at the joint energy distribution of the two photoelectrons obtained using 500~as XUV Gaussian pulses with peak intensity of $4\cdot 10^{10}$~W/cm$^2$ and variable carrier frequency $\omega_0$. 

\begin{figure}[b!]
\includegraphics[width=\columnwidth]{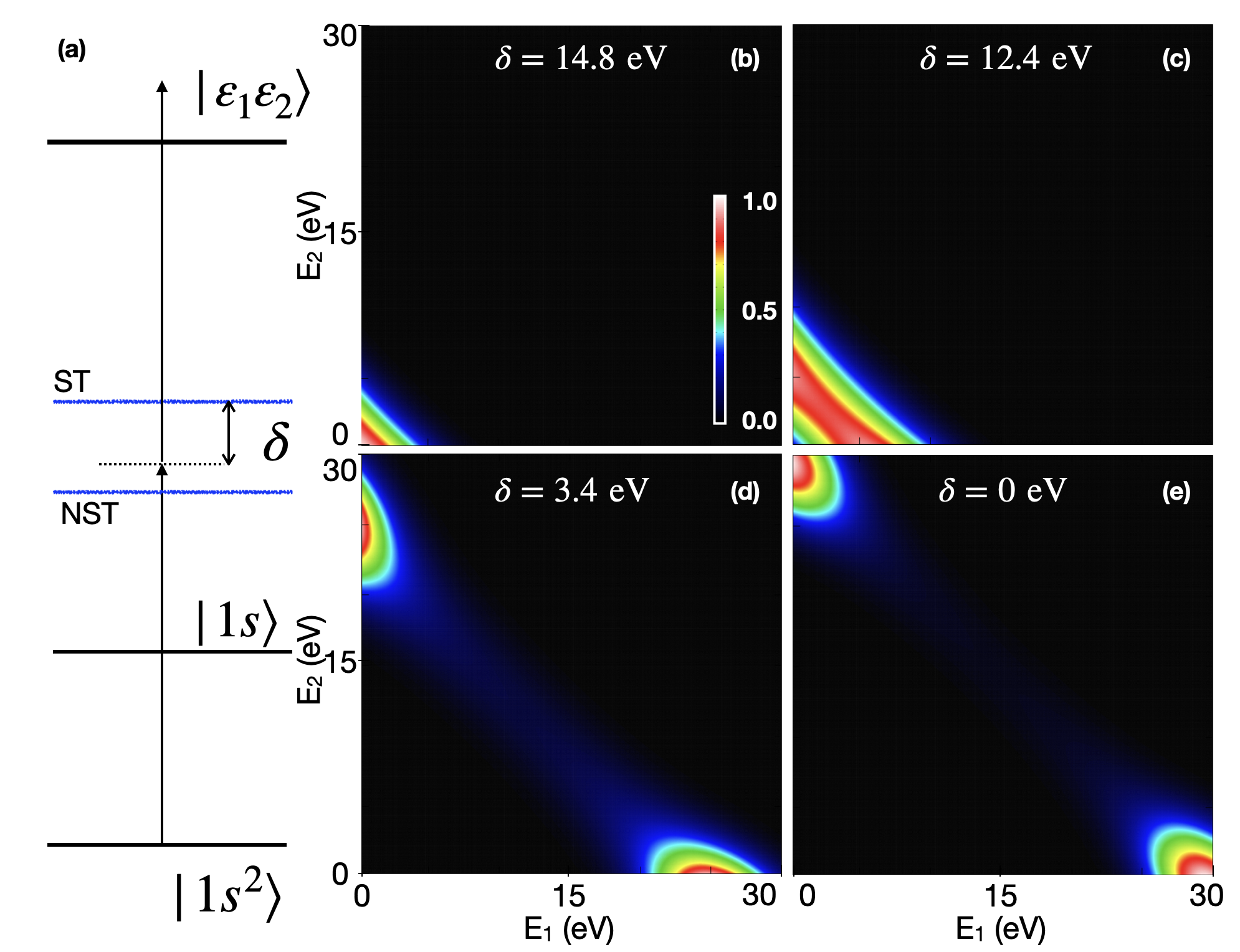}%
\caption{(a) TPDI scheme in the non-sequential regime. $\delta$ is the energy difference from the sequential threshold (ST). Joint energy distribution of two photoelectrons in the non-sequential regime with a fixed pulse duration of 500 as and intensity of $4\cdot10^{10}$~W/cm$^2$ for different central photon energies:  (b) 39.5 eV, (c) 42 eV, (d) 51 eV, and (e) 54.4 eV. The joint energy distributions are normalized by a same factor. In the conditions of the present calculation, a peak signal of $1$ corresponds to a TPDI cross section of  $6\times 10^{-54}$ cm$^4\,\cdot\,$s$\,\cdot\,$eV$^{-2}$ }.
\label{Fig:JED_NonSeq}
\end{figure}
Figure~\ref{Fig:JED_NonSeq}~(a) shows the TPDI scheme used in the present calculation where $\delta$ is the energy difference from the sequential threshold. Fig.~\ref{Fig:JED_NonSeq}~(b) with $\delta$ = 14.8 eV shows the signal appears at the opening threshold of 39.5 eV. As we increase the central photon energy to 42 eV, with $\delta$ = 12.4 eV [Fig.~\ref{Fig:JED_NonSeq}~(c)], the strongly correlated joint energy distribution becomes prominently visible. As shown in Fig.~\ref{Fig:JED_NonSeq}~(d), as we increase the central photon energy to 51 eV, a two-peak structure emerges, which is the hallmark of the sequential mechanism. At the sequential threshold of 54.4 eV the  joint energy distribution  shows the two photoelectrons emitted sequentially with energies of 30 eV, respectively. 

\begin{figure}[t!]
  \includegraphics[width=\columnwidth]{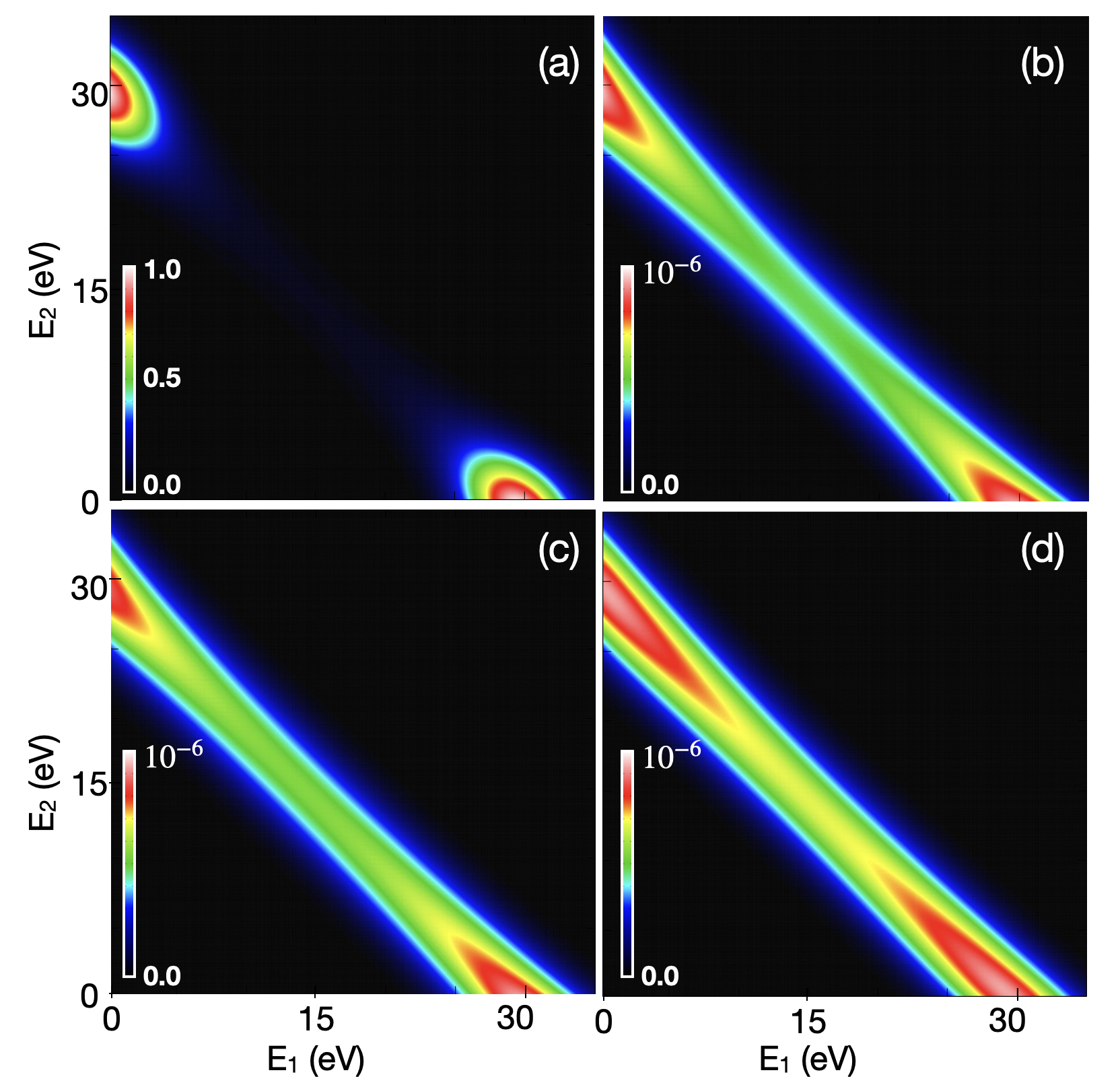}%
\caption{\label{Fig:JED_54eV} Channel-specific contributions to the photoelectron joint energy distribution for TPDI processes promoted by a 500~as $4\cdot10^{10}$~W/cm$^2$pulse with central energy of 54 eV: (a) $\mathrm{1s}\varepsilon\mathrm{p}$; (b) $\mathrm{2p}\varepsilon\mathrm{s}$; (c) $\mathrm{2p}\varepsilon\mathrm{d}$; (d) $\mathrm{2s}\varepsilon\mathrm{p}$. The contribution from the $\mathrm{1s}\varepsilon\mathrm{p}$ intermediate channel dominates. The joint energy distributions are normalized by a same factor. In the conditions of the present calculation, a peak signal of $1$ corresponds to a TPDI cross section of $9\times 10^{-52}$ cm$^4\,\cdot\,$s$\,\cdot\,$eV$^{-2}$}
\end{figure}

To disentangle the contribution from different intermediate ionic states, an independent calculation is performed for each CC channel. Fig.~\ref{Fig:JED_54eV}~(a-d) show the  joint energy distribution  for each of the intermediate states, $\mathrm{1s}\varepsilon\mathrm{p}$, $\mathrm{2s}\varepsilon\mathrm{p}$, $\mathrm{2p}\varepsilon\mathrm{s}$, and $2p\varepsilon\mathrm{d}$, respectively.  With the central photon energy close to the sequential threshold but still well below the $N = 2$ shakeup threshold, virtually all the contribution comes from the dominant $\mathrm{1s}\varepsilon\mathrm{p}$ intermediate state. 

Figure~\ref{Fig:JED_54eV} shows the contribution of individual intermediate channels to the joint energy distribution at $\omega=54$~eV, close to the sequential threshold. At this energy, the distribution resembles that of the sequential mechanism and is dominated by the $1s\epsilon_p$ intermediate channel. At this energy, the $2\ell\epsilon_{\ell'}$ shakeup channels are still closed and hence they contribute only as virtual excitations, which explains the broad distribution of their energy sharing. Indeed, for closed channels, the pole in the denominator of~\eqref{eq:tpdia} does not play any fundamental role, thus leading to a continuum distribution for the energy of the first photoelectron.

\begin{figure}[!htb]
  \includegraphics[width=\columnwidth]{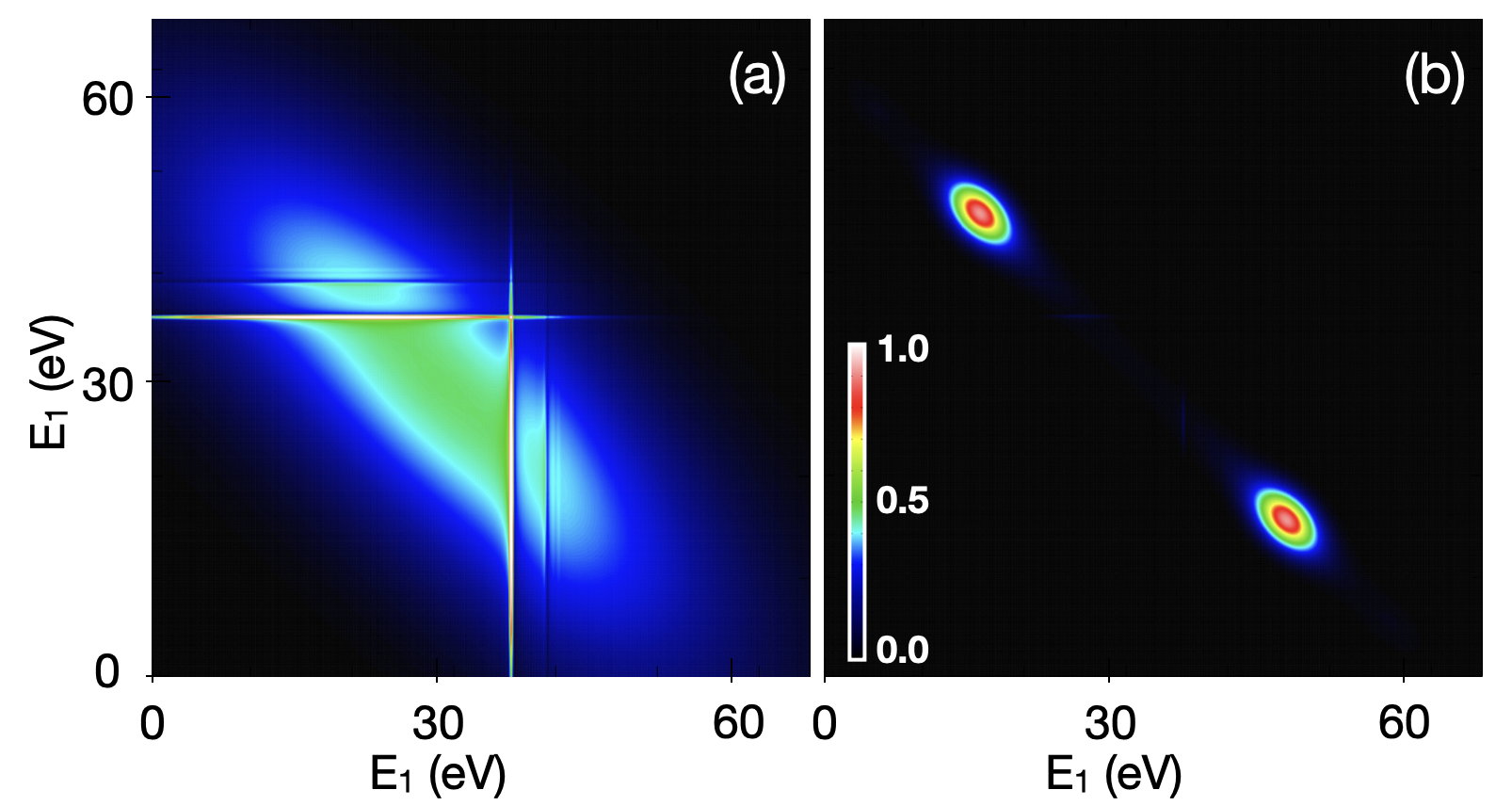}%
  \caption{\label{Fig:JED_70eV} Photoelectron joint energy distribution for TPDI processes promoted by pulses with central energy of 70~eV, peak intensity of $4\cdot10^{10}$~W/cm$^2$ and duration of 120~as (a) and 720~as (b).  The joint energy distributions are normalized by a same factor. In the conditions of the present calculation, a peak signal of $1$ corresponds to a TPDI cross section of $0.5\times 10^{-52}$ cm$^4\,\cdot\,$s$\,\cdot\,$eV$^{-2}$}
\end{figure}

\begin{figure}[!htb]
\includegraphics[width=\columnwidth]{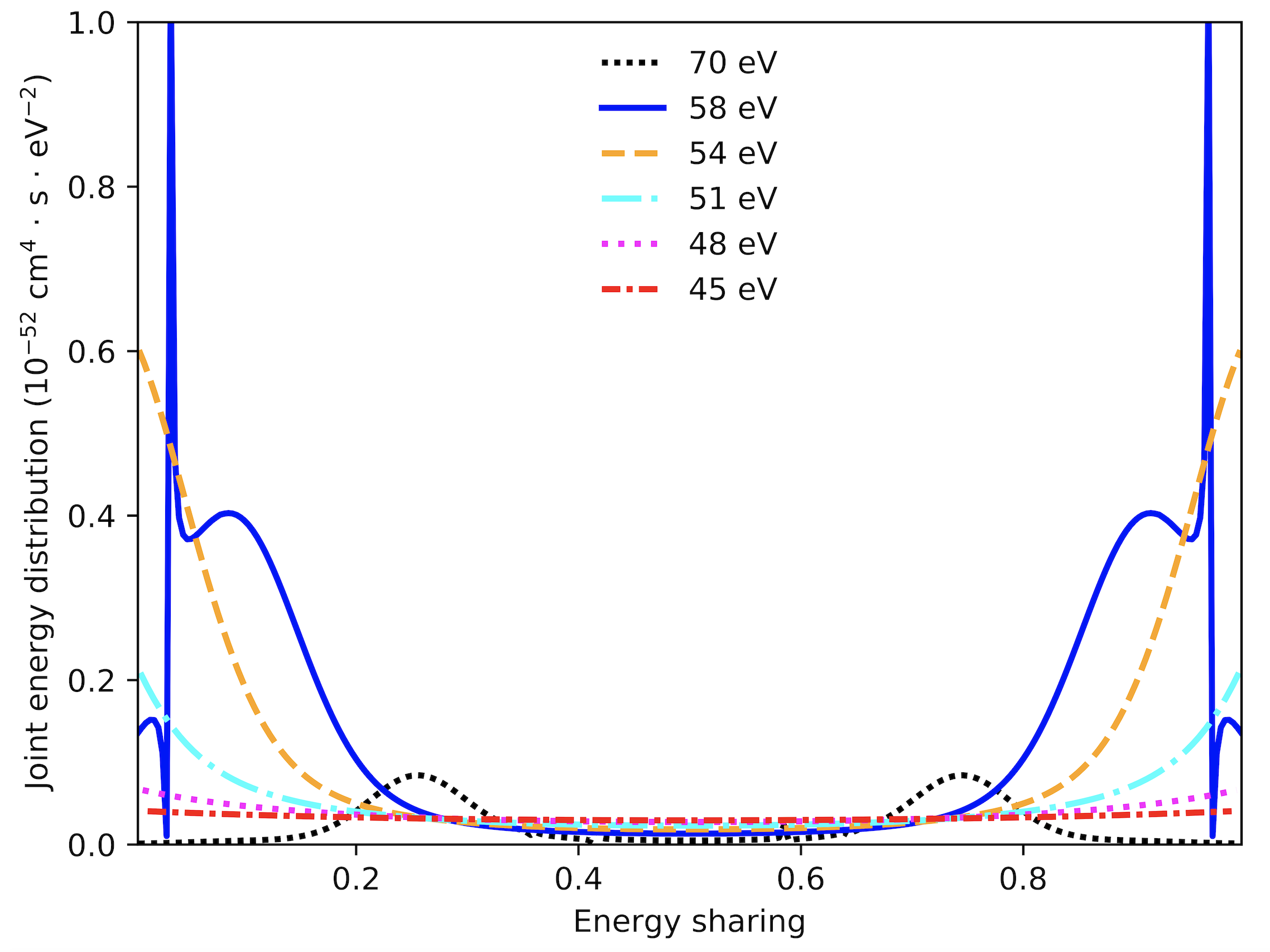}%
\caption{\label{Fig:JED_EnSharing} Joint energy distribution as a function of the photoelectron energy sharing, at different photon energies, from the non-sequential up to the sequential regime. The joint energy distributions are normalized by a same factor so that the peak signal correspond to $1.0 \times 10^{-52}$ cm$^4\,\cdot\,$s$\,\cdot\,$eV$^{-2}$}.
\end{figure}

For ultrashort pulses with central energy above the sequential threshold, the sequential and non-sequential regimes can no longer be separated~\cite{Feist-PRL-09}.
Figure~\ref{Fig:JED_70eV}~(a,b) shows the predictions of the FPVSM for a pulse with central energy of 70~eV and duration of either 120~as (a) or 720~as (b). These plots qualitatively reproduce the results obtained for similar pulses in~\cite{Feist-PRL-09}. In the case of the shorter pulse, in Fig.~\ref{Fig:JED_70eV}~(a), the parent ion does not have the time to relax and hence the joint energy distribution shows a single peak with a strongly correlated distribution. When the longer pulse is employed, on the other hand, the two-peak structure characteristic of the dominant sequential mechanism clearly emerges. Notice that, in the case of the pulse with short duration, the joint energy distribution exhibits also several sharp features. These features are the imprint of the autoionizing states close to the N=2 threshold, which should not be observed and indeed are not reproduced in fully~\emph{ab initio} simulations. As commented in Sec.~\ref{sec:Theo}, the presence of resonant profiles is inherent to the FPVSM, since the ion product of an autoionizing state decay is assumed to be immediately available for ionization. This assumption, however, is obviously not satisfied when the duration of the ionizing pulse is much shorter than the lifetime of the autoionizing states in question, which is the case here. For pulses with duration much longer than an autoionizing state lifetime, on the other hand, the model does capture the contribution of the autoionizing state to the two-photon double ionization due to the resonant enhancement (or suppression) of the bound-continuum ionization amplitude, as discussed in the next subsection. 
Figure~\ref{Fig:JED_EnSharing} shows the photoelectron-pair distribution as a function of the energy sharing $\alpha=E_1/(E_1+E_2)$, generated by 500~as pulses with central energy $\omega_0$ ranging from 45~eV to 70~eV and with peak intensity of $4\cdot 10^{10}$W/cm$^2$. In each case, the distribution is evaluated at the nominal peak of the total energy, $E_1+E_2=2\omega_0-IP$, where $IP$ is the double-ionization potential. In the non-sequential regime, the distribution is almost flat. For energies above the sequential threshold, close to the shake-up threshold, the resonant structure of the autoionizing states emerge. The sharp peak visible for $\omega_0=58$~eV corresponds to the $sp_2^+$ DES. As mentioned earlier, for short pulses, such resonant features are artifacts of the FPVSM. As we approach the double-ionization threshold, the sequential two-peak structure with total energy of 61 eV dominates. 

\subsection{Pump-probe TPDI}
\begin{figure}[!hbtp]
\includegraphics[width=\columnwidth]{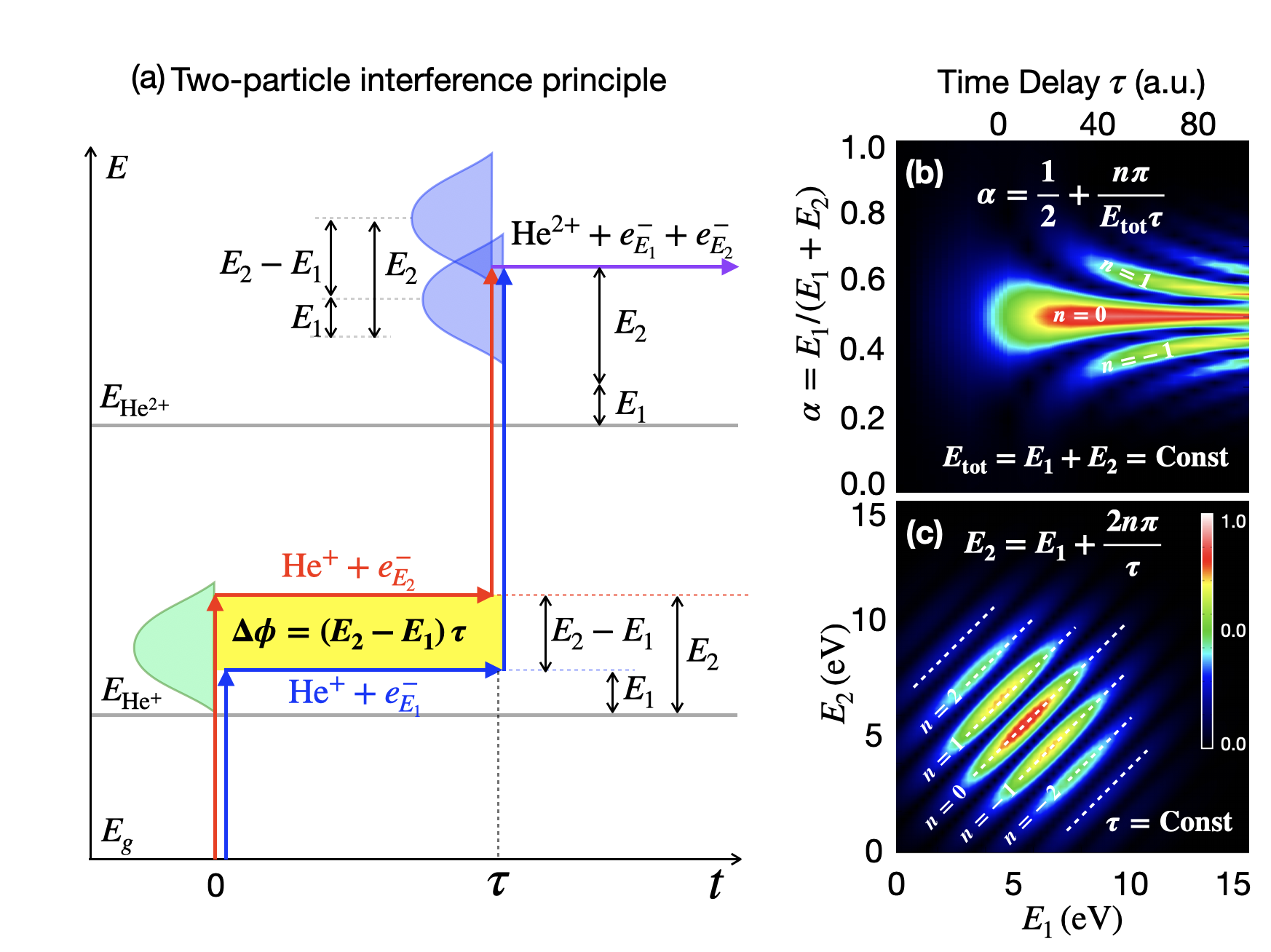}%
\caption{\label{Fig:PumpProbeTPDI} (a) Two-electron quantum interferometric scheme: At time $t$=0, a first XUV photon ionizes the neutral helium atom. After a delay $\tau$, a second XUV photon ionizes $\mathrm{He}^{+}$, creating the $\mathrm{He}^{2+}$ ion and two photoelectrons with energies E$_1$ and E$_2$. Two alternative paths (red and blue, online) contribute to this process, leading to interference fringes in the joint energy distribution. (b)  joint energy distribution as a function of the photoelectron energy sharing and the pump-probe delay. (c) joint energy distribution as a function of two photoelectron energies for a specific time-delay, $\tau=2$~fs.}
\end{figure}

Two-particle quantum interference can be used to probe the entanglement between two identical particles~\cite{Horne-PRL-89}. In the context of photoionization studies, McCurdy and collaborators have shown the two-particle interference in the joint energy distribution of the TPDI process in helium~\cite{Palacios-PRL-09, Palacios-JPB-10}. For the present case, we consider a pump-probe scheme of the TPDI of helium with two XUV pulses with a controllable delay $\tau$,
\begin{equation}
    \vec{\mathcal{E}}(t) = \vec{\mathcal{E}}_{\textsc{XUV}_1}(t) + \vec{\mathcal{E}}_{\textsc{XUV}_2}(t-\tau),
\end{equation}
where $\vec{\mathcal{E}}_{\textsc{XUV}_{1/2}}(t)$ indicate the transverse electric fields of the two pulses. In the present calculation, the two pulses have central energies $\omega_1=30$~eV and $\omega_2=60$~eV and duration of 1~fs. 
At negative time delays, when the 60~eV XUV pulse comes first, the 30~eV XUV pulse is unable to ionize the residual $\mathrm{He}^{+}(1s)$ ion, which has ionization potential 54.4 eV, leading to no TPDI signal around $E_1+E_2\simeq 11$~eV. If the more energetic pulse exceeded the $2s/2p$ shake-up threshold, one could have observed a signal, since the ionization potential of the excited $\mathrm{He}^+$ ion is below 30~eV.
At positive time delays, the 30~eV pump pulse ionizes the neutral helium atom and, at some later time $\tau$, the probe pulse ionizes the $\mathrm{He}^{+}$ ion leading to $\mathrm{He}^{2+}$ and two photoelectrons with energies $E_1$ and $E_2$, as illustrated in Fig.~\ref{Fig:PumpProbeTPDI}~(a). 
Thanks to the finite spectral width of the two pulses, the final state can be reached through two distinct paths. In one path (blue arrows in the figure), the first ionization event generates a photoelectron with energy $E_1$ through the absorption of a photon on the lower-frequency edge of the pump-pulse spectrum, whereas in the second ionization event the ion absorbs a photon with frequency on the upper end of the probe-pulse spectrum. In the other path (red arrows in the figure), the order with which the two photoelectrons are generated is reversed. In the interval between the two pulses, the energy of the system along the two path differ by $\Delta E = E_2-E_1$, and hence the two paths acquire a phase difference $\Delta \phi=(E_2-E_1)\tau$. Since the final state is the same for the two paths, the associated amplitudes interfere constructively or destructively if $\Delta \phi$ is an even or odd integer multiple of $\pi$, respectively. From the energetic point of view, this interference scheme is analogous to the Ramsey interference observed in attosecond pump-probe single-ionization processes~\cite{Mauritsson2010,Argenti2010}, where the first step excites the system to different bound metastable states. Here, both events eject a particle and the interference occurs because the two particles being ejected are identical. Constructive interference is realized for $\Delta \phi=2n\pi$,
\begin{equation}
    E_2=E_1+\frac{2\,n\,\pi}{\tau}, \qquad n\in\mathbb{Z}.
\end{equation}
In the photoelectron joint energy distribution, therefore, the interference fringes appear as straight lines parallel to the $E_2=E_1$ diagonal, and separated by a distance $d=\sqrt{2}\pi/\tau$ as shown in Fig.~\ref{Fig:PumpProbeTPDI}~(c), as it was already highlighted in~\cite{Palacios-PRL-09, Palacios-JPB-10}. Interference fringes are visible also in the energy-sharing spectrum, at a fixed total photoelectron energy $E_{\textrm{tot}}=E_1+E_2$, as a function of the delay, in which case the condition for constructive interference becomes
\begin{equation}
    \alpha = \frac{1}{2}+\frac{n\,\pi}{E_{\mathrm{tot}}\,\tau},\qquad n\in\mathbb{Z}.
\end{equation}
In this case, therefore, the fringes have hyperbolic profiles, as shown in Fig.~\ref{Fig:PumpProbeTPDI}~(b), reminescent of those observed in single-ionization attosecond photoelectron spectra~\cite{Mauritsson2010}. As the time delay between the pulses increases, of course, the interference fringes become difficult to resolve experimentally, and the total signal converges to the incoherent sum of those for the two alternative paths.

As a final example, we examine the FPVSM predictions for the resonant TPDI of helium enhanced by an intermediate DES, using a two-color XUV pump-probe scheme with duration longer than the lifetime of the DES. In all the cases considered in the following, the two pulses have zero relative delay.
\begin{figure}[b!]
  \includegraphics[width=\columnwidth]{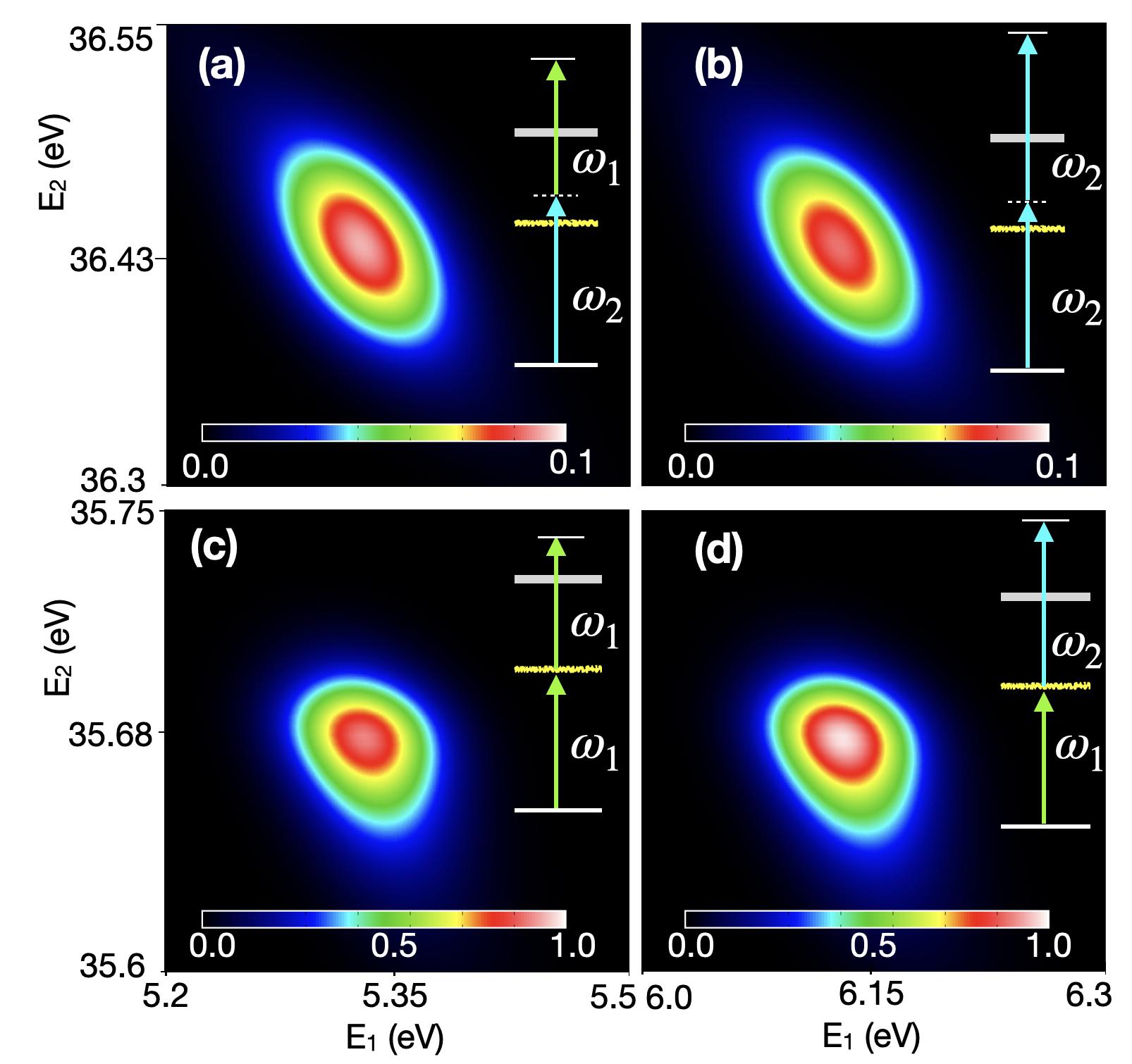}%
  \caption{\label{Fig:PumpProbeTPDI_40fs} Two-color TPDI: joint energy distribution, as a function of two photoelectron energies, generated by the absorption of two XUV photon from 40~fs pulses with central energies $\omega_1=$ 60.15~eV and $\omega_2=$ 60.9~eV, and zero relative delay. The lower frequency, $\omega_1$, is in resonance with the $\mathrm{sp_{2}^{+}}$ bright doubly-excited state, which is 5.04 eV below the N~=~2 threshold and has a lifetime of 17.6 fs. In each panel, the pathway corresponding to the associated signal is shown. The joint energy distributions are normalized by a same factor. In the conditions of the present calculation, a peak signal of $1$ corresponds to a TPDI cross section of $1.5\times 10^{-51}$ cm$^4\,\cdot\,$s$\,\cdot\,$eV$^{-2}$.}
\end{figure}
In the past, the role of DES in TPDI of $\mathrm{He}$ has been studied by solving the TDSE with ultrashort pulses~\cite{Feist-PRL-11}. So far, however, TDSE-based approaches have considered only intense XUV pulses with attosecond or few femtosecond duration, which are difficult to realize experimentally. Here we probe these resonances with longer pulses, which are more easily produced at XFELs facilities. As a case study, we select the $\mathrm{sp_{2}^{+}}$ DES intermediate state, which has a lifetime of 17.6~fs~\cite{Rost-JPB-97}. We study the effect of this state on the joint photoelectron energy distribution in the TPDI of helium with a pair of XUV pulses, each with 40~fs duration and intensity $I=10^{10}\,\mathrm{W/cm^2}$. 

Figure~\ref{Fig:PumpProbeTPDI_40fs} shows the joint energy distribution when the two pulses have central energy of 60.15~eV and 60.90~eV. The first pulse is resonant with the $1s^2-sp_{2}^+$ transition, $E_{sp_2^+}-E_g=60.15$~eV. The distribution exhibits four distinct peaks in the $E_2>E_1$ portion of the spectrum, corresponding to whether both photons are provided from the first pulse [Fig.~\ref{Fig:PumpProbeTPDI_40fs}\,(c)], from the second pulse [Fig.~\ref{Fig:PumpProbeTPDI_40fs}\,(b)], or one from each pulse [Fig.~\ref{Fig:PumpProbeTPDI_40fs}\,(a,d)]. These signals correspond to two resonant (c,d) and two non-resonant (a,b) transitions. Since the peaks are narrow compared with their energy distance in the 2D spectrum, we show them magnified in four separate panels for clarity. The peaks for the non-resonant transitions are symmetric relative to the equal total energy axis, as observed already in the previous section, with ultrashort pulses. Remarkably, the energy distribution is not symmetric anymore in the two resonant cases, which is to be expected given the strong modulation of the first resonant one-photon transition amplitude. In the FPVSM, the resonant modulation of the signal follows the Fano profile of the resonance. Indeed, according to~\eqref{eq:2PDIAmp}, the total ionization amplitude is a linear combination of products of dipole transition amplitudes. With well-separated signals in the joint spectrum, one of these products dominates, owing to the field factor. In the resonant case, therefore, the amplitude factorises into the product of a structureless field form factor times a structureless hydrogenic ionization amplitude of the residual ion and a resonant ionization amplitude from the ground state to the two-electron continuum.

\begin{figure}[!htb]
  \includegraphics[width=\columnwidth]{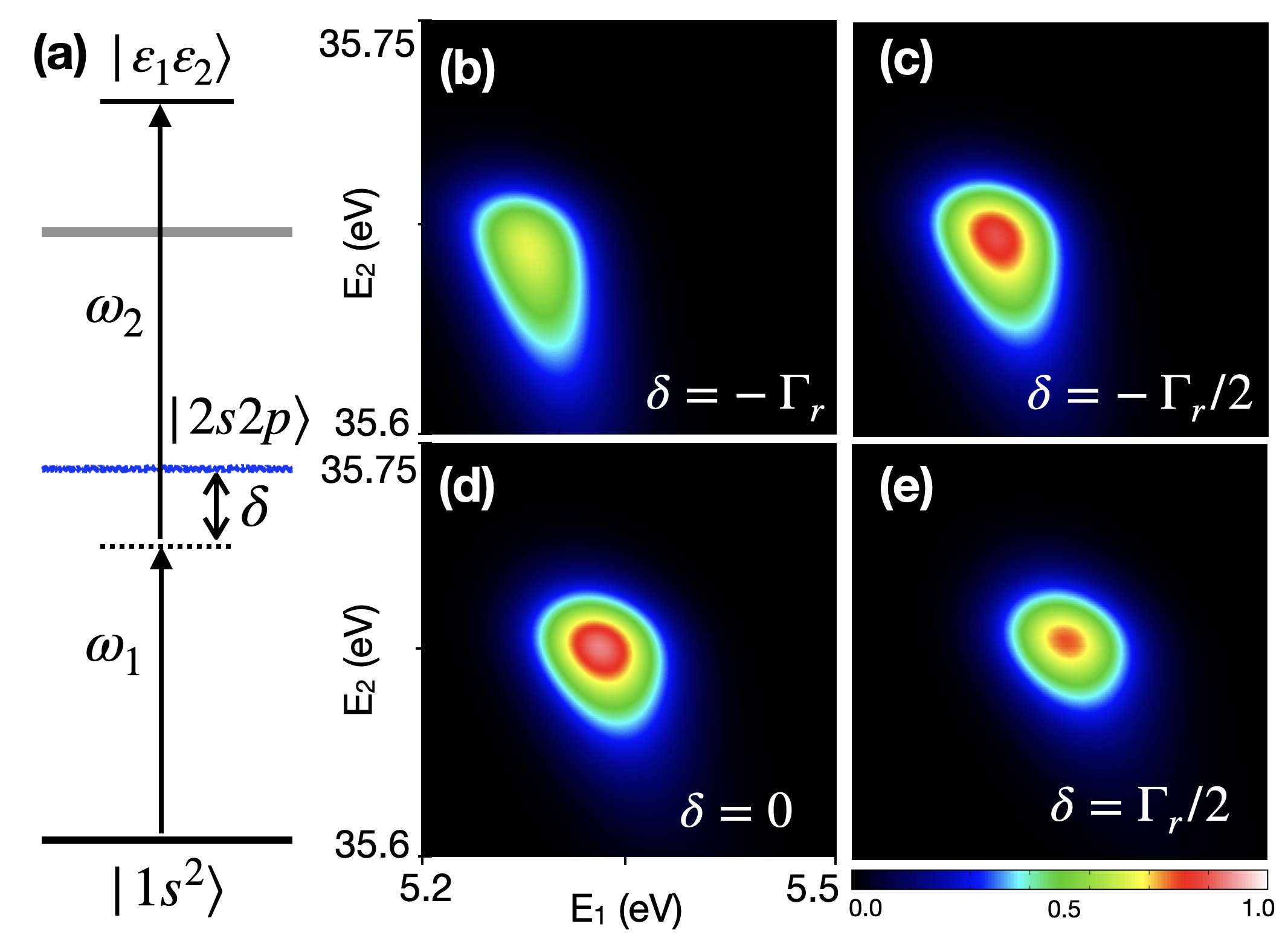}%
  \caption{(a) Schematic of the resonant TPDI process, which strongly depends on the detuning $\delta$. (b-e) Joint energy distribution for four different calculations of the resonant TPDI process with fixed $\omega_2=60.9$~eV and variable $\omega_1=E_{sp_2^+}-E_g+\delta$, where the detuning is indicated in the panels in units of the resonance width $\Gamma_r=0.037$~eV. Both pulses have a duration of 40~fs, intensity of $10^{10}$W/cm$^2$, and zero relative delay. The joint energy distributions are normalized by a same factor. In the conditions of the present calculation, a peak signal of $1$ corresponds to a TPDI cross section of $1.5\times 10^{-51}$ cm$^4\,\cdot\,$s$\,\cdot\,$eV$^{-2}$.}
  \label{Fig:PumpProbeTPDI_delta}
\end{figure}
Figure~\ref{Fig:PumpProbeTPDI_delta}~(b-e) show the resonant peak for the transition ~\ref{Fig:PumpProbeTPDI_delta}~(a), for four different values of the detuning $\delta$ of the pump-pulse central energy from the $sp_2^+$ resonance. The probe photon has a fixed central frequency $\omega_2=60.9$~eV. The $sp_2^+$ has a Fano profile with a negative $q$ parameter, i.e., the transition amplitude increases monotonically with energy, peaks, and drops to zero at an energy above the resonant peak, before slowly returning to the background value. This is reflected in the shape of the profile in the four panels: at negative detuning, the signal is stretched below the resonant peak (tail, panel b, c). At positive detuning, the peak and adjacent zero are illuminated, leading to a signal with energy breath considerably narrower than the pump pulse's and comparable to the resonant width. 
Even if the resonant profiles of these results are predictable, they are a valuable starting point to assess the deviation of the signal, in the exact resonant double ionization case, due to the partial decay of a resonance, or the contribution of its direct double ionization by the probe pulse, for helium as well as for more complex atoms.

\section{\label{sec:con}Conclusion}
In this work, we introduced a finite pulse version of the virtual sequential model (FPVSM), based on \emph{ab initio} multi-channel one-photon transition amplitudes, to describe the two-photon double ionization process in polyelectronic atoms and, as a proof of principle, we used it to reproduce several features of results for the TPDI of atomic helium, obtained with TDSE simulations. We have calculated the joint energy distribution of the two photoelectrons and the energy sharing from the non-sequential to the sequential regime. The FPVSM allows us to compute the strongly correlated photoelectron joint energy distribution in the non-sequential regime, and the uncorrelated counterpart in the sequential regime, more efficiently than a full TDSE simulation. Unlike previous TDSE simulations, the CC approach in FPVSM allows us to quantify the contribution from different channels and  highlights the features associated with each intermediate state under consideration. Furthermore, the model captures how the transition from the non-sequential to sequential blurs as one considers extreme ultrashort pulses with energies near the double ionization threshold. We demonstrate that the energy sharing between the photoelectrons significantly changes as we approach the double ionization threshold and how the two-peak structure emerges in the sequential regime. The model is capable of reproducing the salient features of two-particle interference, already observed in TDSE simulations, which highlight its explanatory power. We have also applied the model to study the asymmetry of the resonant TPDI photoelectron joint energy distribution close to the optically allowed $\mathrm{sp_{2}^{+}}$ doubly-excited state with long XUV-pulses, which is a regime not explored by TDSE simulations. The model admits a natural extension to polyelectronic atoms, which present the additional interesting feature of multiple grand-parent ions. An application of this model to complex atoms such as neon and argon is ongoing and will be subject of future work.

\begin{acknowledgments}
 This work is supported by NSF grant No. PHY-1912507 and by the DOE CAREER grant No. DE-SC0020311.  We express our gratitude to Jeppe Olsen for many useful discussions.
\end{acknowledgments}

\appendix
\section{\label{app:Recoupling}Two-photon double ionization matrix element}

This appendix details the derivation of the TPDI matrix element~\eqref{eq:2PDIAmp}. Grandparent-ion wavefunctions ($N-2$ electrons) are denoted by the symbol $\Phi_A$, and parent-ion wavefunctions ($N-1$ electrons) by the symbol $\Phi_a$. 
The $N-$electron states we will consider here are either bound, $|\Psi_n\rangle$, single-ionization, or double ionization states. 
In the calculation of the $\langle A\,\vec{k}_{1}\sigma_1\vec{k}_{2}\sigma_2| \mathcal{O}_{\nu}|\Psi_{\alpha \vec{k}^{\prime}}^-\rangle$,
the single-ionization wave-functions are approximated using the single-channel functions, 
\begin{equation}\label{eq:PsiMinus}
\Psi^{-}_{\alpha\vec{k}^{\prime} }\simeq
\sqrt {N}\hat{\mathcal{A}}_{N}\,\Xi_{\alpha}^\Gamma(x_1\longdash x_{N-1};\hat{r}_N,\zeta_N)\,\phi_{\alpha,\ell_\alpha \epsilon}^{\Gamma\,-}(r_N)
\end{equation}
where in $\Xi_{\alpha}^\Gamma(x_1\longdash x_{N-1};\hat{r}_N,\zeta_N)$ the parent ion is coupled to the angular and spin part of the $N$-th electron to give rise to a well-defined angular momentum and spin, which are specified in the collective total-symmetry index $\Gamma=(S,L,\Pi;\Sigma,M)$,
\begin{eqnarray}
&&\Xi_{\alpha}^\Gamma(x_1\longdash x_{N-1};\hat{r}_N,\zeta_N) =\nonumber\\
&&=[[\Phi_{a_\alpha}(x_1\longdash x_{N-1})\otimes Y_{\ell_\alpha}(\hat{r}_N)]_{LM}\otimes {^2\chi(\zeta_N)}]_{S\Sigma}= \nonumber\\
&&=\sum_{M_a m}\sum_{\Sigma_a \sigma} 
C_{L_a M_a,\ell_\alpha m}^{LM}
C_{S_a \Sigma_a,\frac{1}{2}\sigma}^{S\Sigma}\times\nonumber\\
&&\times
\Phi_{a_\alpha,M_a \Sigma_a}(x_1\longdash x_{N-1})Y_{\ell_\alpha m}(\hat{r}_N) {^2\chi_\sigma(\zeta_N)}.
\end{eqnarray}
In~\eqref{eq:PsiMinus}, the radial photoelectron wavefunction $\phi_{\alpha,\ell_\alpha \epsilon}^{\Gamma\,-}(r)$ is normalized in such a way that its outgoing component is
\begin{equation}
\left[\phi_{\alpha,\ell_\alpha \epsilon}^{\Gamma\,-}(r)\right]_{\mathrm{out}} = \frac{1}{\sqrt{2\pi k_\alpha}}\frac{e^{i\theta_\alpha(r)}}{r}, \qquad r\to\infty
\end{equation}
with $k_\alpha = \sqrt{2\epsilon}$, and $\theta_\alpha(r) = k_\alpha r +\frac{Z}{k_\alpha} \ln 2k_\alpha r - \ell\pi/2 + \sigma_{\ell_\alpha}(k_\alpha)$ is the Coulomb phase factor. Finally, $\mathcal{A}_N$ is the idempotent antisymmetrizer,
\begin{equation}
\mathcal{A}_N=\frac{1}{N!}\sum_{\mathcal{P}\in\mathcal{S}_N}(-1)^{\mathrm{sgn} \mathcal{P}}\mathcal{P},\qquad
\mathcal{A}_N^2=\mathcal{A}_N=\mathcal{A}_N^\dagger.
\end{equation}

To define a double-ionization channel, we need to specify, beyond the total quantum numbers $\Gamma$ = $(S,L,\Pi;\Sigma,M)$, the state of the grandparent ion, $\Phi_{A}(x_1\longdash x_{N-2})$ and the non-energy quantum numbers of the two free electrons, namely, their orbital angular momenta, $\ell_1$, and $\ell_2$, and the angular coupling scheme. Asymptotically
\begin{eqnarray}
\Psi^-&&\phantom{}_{A, E_1 \ell_1 m_1 \sigma_1, E_2 \ell_2 m_2 \sigma_2} =\sqrt{N(N-1)}\hat{\mathcal{A}}_N\,\times\nonumber\\
&&\times\Psi_A(x_1\longdash x_{N-2})\otimes\nonumber\\
&&\otimes{^{2}\phi^-_{E_1\ell_1 m_1\sigma_1}}(x_{N-1})\otimes {^{2}\phi^-_{E_2\ell_2 m_2\sigma_2}}(x_N).\label{eq:Psi-l}
\end{eqnarray}
This expression can be cast in a symmetrized combination in which either electron $1$ or $2$ is recoupled with the grand-parent-ion $A$ to give rise to a scattering state for the system with $N-1$ electron. Once the transition amplitudes to the states in~\eqref{eq:Psi-l} are known, the fully differential TPDI amplitude is readily reconstructed using the expansion of energy-normalized Coulomb plane waves,
\begin{equation}
\begin{split}
\psi_{E\hat{\Omega}}^-(\vec{r})&=\sqrt{\frac{2k}{\pi}}\sum_{\ell m} i^{\ell}e^{-i\sigma_\ell} F_\ell(kr)Y_{\ell m}^*(\hat{k})Y_{\ell m}(\hat{r})\\
&=\sum_{\ell m} i^{\ell-1}e^{-i\sigma_\ell}Y_{\ell m}^*(\hat{k}) \phi^-_{\ell m E}(\vec{r}).
\end{split}
\end{equation}

As the next step in the application of the FPVSM, one of the two photoelectron is angularly and spin and permutationally coupled to the grand-parent ion, using the well-known identities $\delta_{\alpha\alpha'}\delta_{\beta\beta'}=\sum_{c\gamma} C_{a\alpha,b\beta}^{c\gamma}C_{a\alpha',b\beta'}^{c\gamma}$ and $\hat{\mathcal{A}}_N = \frac{1}{N}[1-(N-1)\mathcal{P}_{N-1,N}]\hat{\mathcal{A}}_{N-1}$. Finally, the recoupled grand-parent/photoelectron state is identified with the scattering state of the singly ionized system that satisfies the same outgoing boundary conditions. Since this identification is in itself an approximation, it breaks the symmetry between the two photoelectrons. To avoid such bias in the result, therefore, it is convenient to symmetrically split~\eqref{eq:Psi-l} first and, in each of the two resulting identical components, couple the grandparent ion to either the first or the second photoelectron. The result of this tedious but straightforward process is
\begin{widetext}
\begin{equation}\label{eq:dich3}
\begin{split}
\Psi^-_{A, E_1 \ell_1 m_1 \sigma_1, E_2 \ell_2 m_2 \sigma_2}
&=\frac{1-(N-1)\mathcal{P}_{N-1,N}}{2\sqrt{N}}
\sum_{\Gamma_\aleph} 
C_{L_A M_A, \ell_1 m_1}^{L_\aleph M_\aleph} 
C_{S_A \Sigma_A, \frac{1}{2}\sigma_1}^{S_\aleph \Sigma_\aleph}
\Psi^-_{\aleph E_1}(x_1\longdash x_{N-1})\otimes {^{2}\phi^-_{E_2\ell_2 m_2\sigma_2}}(x_N)-\\
&-\frac{1-(N-1)\mathcal{P}_{N-1,N}}{2\sqrt{N}}
\sum_{\Gamma_\beth} 
C_{L_A M_A, \ell_2 m_2}^{L_\beth M_\beth} 
C_{S_A \Sigma_A, \frac{1}{2}\sigma_2}^{S_\beth \Sigma_\beth}
\Psi^-_{\beth E_2}(x_1\longdash x_{N-1})\otimes {^{2}\phi^-_{E_1\ell_1 m_1\sigma_1}}(x_N)
\end{split}
\end{equation}
where the summations are constrained so that $\ell_\aleph = \ell_1$, $\ell_\beth=\ell_2$, and we have introduced 
\begin{equation}
\Psi^-_{\aleph E_1}(x_1\longdash x_{N-1}) = \sqrt{N-1}\,\hat{\mathcal{A}}_{N-1}\sum_{M_A' m_1' \Sigma_A' \sigma_1'} \hspace{-12pt}C_{L_A M_A', \ell_1 m_1'}^{L_\aleph M_\aleph}
C_{S_A \Sigma_A', \frac{1}{2}\sigma_1'}^{S_\aleph \Sigma_\aleph}\Psi_{A,\Sigma_A',M_A'}(x_1\longdash x_{N-2})\otimes {^{2}\phi^-_{E_1\ell_1 m_1'\sigma_1'}}(x_{N-1}).
\end{equation}
\end{widetext}
To evaluate the dipole matrix element between a double ionization and a single ionization continuum, we further assume that the unbound electron in the latter does not participate in the transition and plays the role of a spectator instead. 

Using Wigner-Eckart theorem and the orthogonality of the continuum wave functions, we get
\begin{widetext}
\begin{equation}\label{eq:single2doubledipole}
\begin{split}
&\langle \Psi^-_{A, E_1 \ell_1 m_1 \sigma_1, E_2 \ell_2 m_2 \sigma_2} | \mathcal{O}_{\nu} |
\Psi_{\alpha E}^{\Gamma(-)}\rangle =\\
=&\frac{1}{2}
\sum_{\Sigma_\aleph}
C_{S_A \Sigma_A, \frac{1}{2}\sigma_1}^{S_a \Sigma_\aleph}
C_{S_a \Sigma_\aleph,\frac{1}{2}\sigma_2}^{S\Sigma}
\sum_{L_\aleph}
\frac{1}{\Pi_{L_\aleph}}
\langle \Psi^-_{\aleph E_1}\|\mathcal{O}_1\|
\Phi_{a_\alpha}\rangle\,\delta_{\ell_2 \ell_\alpha} 
\delta(E-E_a-E_2)
\sum_{M_\aleph M_a}
C_{L_A M_A, \ell_1 m_1}^{L_\aleph M_\aleph} 
C_{L_a M_a,\ell_2 m_2}^{LM}
C_{L_a M_a, 1 \nu}^{L_\aleph M_\aleph}
-\\
-&\frac{1}{2}
\sum_{\Sigma_\beth}
C_{S_A \Sigma_A, \frac{1}{2}\sigma_2}^{S_a \Sigma_\beth}
C_{S_a \Sigma_\beth,\frac{1}{2}\sigma_1}^{S\Sigma}
\sum_{L_\beth}
\frac{1}{\Pi_{L_\beth}}
\langle \Psi^-_{\beth E_2} \|\mathcal{O}_{1} \|
\Phi_{a_\alpha}
\rangle\,
\delta_{\ell_1 \ell_\alpha} \delta(E-E_a-E_1)
\sum_{M_\beth M_a}
C_{L_A M_A, \ell_2 m_2}^{L_\beth M_\beth} 
C_{L_a M_a,\ell_1 m_1}^{LM}
C_{L_a M_a, 1 \nu}^{L_\beth M_\beth}
\end{split}
\end{equation}
Now, we can evaluate the two-photon matrix element
\begin{equation}
\begin{split}
&\langle \Psi^-_{A, E_1 \ell_1 m_1 \sigma_1, E_2 \ell_2 m_2 \sigma_2} | \mathcal{O}_{\nu}G_0^+(E_g+\omega)\mathcal{O}_{\mu}|g\rangle=
\sum_{\Gamma\alpha} \int dE \frac{\langle \Psi^-_{A, E_1 \ell_1 m_1 \sigma_1, E_2 \ell_2 m_2 \sigma_2} | \mathcal{O}_{\nu}|\Psi_{\alpha E}^{\Gamma(-)}\rangle\,\langle\Psi_{\alpha E}^{\Gamma(-)}|\mathcal{O}_{\mu}|g\rangle}{E_g+\omega-E+i0^+}
\end{split}
\end{equation}
In the relevant special case of an initial states with {$^1$S} symmetry,
\begin{equation}
\begin{split}
&\langle \Psi^-_{A, E_1 \ell_1 m_1 \sigma_1, E_2 \ell_2 m_2 \sigma_2} | \mathcal{O}_{\nu}G_0^+(E_g+\omega)\mathcal{O}_{\mu}|g\rangle=\\
=&
\frac{1-\mathcal{P}_{12}}{2\sqrt{3}} 
\frac{C_{\frac{1}{2}\sigma_2,\frac{1}{2}\sigma_1}^{S_A-\Sigma_A}}{\Pi_{S_A}}
\sum_{a L}
\frac{1}{\Pi_{L}}
\frac{
\langle \Psi^{^{2S_a+1}L^{\bar{\pi}_a}(-)}_{A \ell_1 E_1}\|\mathcal{O}_1\|\Phi_{a}\rangle\,
\langle\Psi_{a\ell_2 E_2}^{{^1P^o}(-)}\|\mathcal{O}_1\|g\rangle
}{E_g+\omega-E_a-E_2+i0^+}
\sum_{M M_a}
C_{L_A M_A, \ell_1 m_1}^{L M} 
C_{L_a M_a,\ell_2 m_2}^{1 \mu}
C_{L_a M_a, 1 \nu}^{L M},
\end{split}
\end{equation}
where $\bar{e}=o$ and vice versa. Convolution with the external field~\eqref{eq:2PT_FrequencyIntegral} readily yields~\eqref{eq:2PDIAmp}.
\end{widetext}

\section{\label{app:FrequencyIntegral} Two-photon frequency integral}

In this appendix, we derive the general formula for the two-photon frequency integral that appears in Eq.~\eqref{eq:2PDIAmp} in terms of  Faddeyeva function,
\begin{equation}
I_{21} = \int d\omega \frac{\tilde{A}_2(E_n-\omega)\tilde{A}_1(\omega)}{E_p+\omega+i0^+},
\end{equation}
where the numerator frequency detuning $E_n$ and the pole shift $E_p$ are known functions of the final energy of the two photoelectrons as well as of the energy of the intermediate and final ion, whereas $\tilde{A}_i(\omega)$ ($i=1,2$) are the spectra of two unchirped Gaussian pulses.
Here, the vector potentials of the two \emph{linearly polarized} pulses in the time domain (notice that since we are considering arbitrary sequences of linearly polarized pulses, we are also automatically including the case of arbitrarily polarized pulses as well), $\vec{A}_i(t)$, are defined as
$\vec{A}_i(t) = \hat{\epsilon}_i A_i e^{-\sigma_i^2(t-t_i)^2/2}\cos\left[(\omega-\omega_i)t+\varphi_i\right]$,
where $\hat{\epsilon}_i$ is the light polarization, $A_i$ is the vector potential amplitude, $\varphi_i$ the carrier-envelope phase, $\omega_i$ the central frequency,  and $\sigma_i$  the standard deviation of the vector-potential spectrum. The temporal duration of the pulse, normally identified with the full-width at half maximum (\textsc{fwhm}) of the pulse intensity, is then \textsc{fwhm}$=2\sqrt{\ln 2}\sigma_i^{-1}$.
In the following it is useful to split each pulse $\vec{A}_i(t)$ in its positive- and negative-central-frequency components $\vec{A}^\pm_i(t)$, 
$\vec{A}_i(t) = \vec{A}^+_i(t) + \vec{A}^-_i(t)$,
$\vec{A}^\pm_i(t) = \hat{\epsilon}_i \frac{A_i}{2} e^{-\sigma_i^2(t-t_i)^2/2}e^{\pm i \left[(\omega-\omega_i)t+ \varphi_i\right]}$.
The pulse spectrum,  defined as
\begin{equation}
\tilde{A}_{i}(\omega)\equiv (2\pi)^{-1/2}\int dt\,\, \hat{\epsilon}_i^*\cdot A_i(t) e^{i\omega t},
\end{equation}
also separates in the sum of a positive- and a negative-central-frequency component, $\tilde{A}_i(\omega)= \tilde{A}_i^+(\omega)+\tilde{A}_i^-(\omega)$,
\begin{equation}
\begin{split}
\tilde{A}_i^\pm(\omega)&= \frac{A_i}{2\sigma_i}\exp\left[i(\omega t_i \mp \varphi_i)-(\omega\mp\omega_i)^2/2\sigma_i^2\right].
\end{split}
\end{equation}
In the case of the absorption of two photons,
\begin{equation}\label{eq:I21pi2pi1}
I_{21} = \int d\omega \frac{\tilde{A}_2^{+}(E_n-\omega)\tilde{A}_1^{+}(\omega)}{E_p+\omega+i0^+}.
\end{equation}
Let us introduce the delay $\tau$ between the two pulses, $\tau=t_2-t_1$. The numerator in~\eqref{eq:I21pi2pi1} then reads
\begin{eqnarray}
&&\tilde{A}_2^{+}(E_n-\omega)\tilde{A}_1^{+}(\omega) = 
\frac{A_1A_2}{4\sigma_1\sigma_2}\,
e^{-i(\varphi_1+\varphi_2)}\,
e^{iE_nt_2}\times\nonumber\\
&\times&\exp\left[- 
\frac{(\omega-\omega_1)^2}{2\sigma_1^2}-
\frac{\left(\omega-E_n+\omega_2\right)^2}{2\sigma_2^2}-  i\omega\tau\right]
\end{eqnarray}
and the frequency integral $I_{21}$ can be rewritten as
\begin{eqnarray}
I_{21}&=& 
\frac{A_1A_2}{4\sigma_1\sigma_2}\,
e^{-i(\varphi_1+\varphi_2-E_nt_2)}\times\nonumber\\
&\times&e^{-\omega_1^2/2\sigma_1^2-(E_n-\omega_2)^2/2\sigma_2^2
+ \sigma_t^2(\tilde{\omega}-E_p)^2/2}\times\nonumber\\
&\times&\int_{-\infty}^{\infty} d\omega\,\frac{\exp[-\sigma_t^2(\omega-\tilde{\omega})^2/2]}{\omega+i0^+},\label{eq:FrequencyIntegral}
\end{eqnarray}
where we have introduced the new parameters
\begin{eqnarray}
\sigma_t &=& \sqrt{1/\sigma_1^2+1/\sigma_2^2},\\
\tilde{\omega} &=& E_p + \left(\frac{\omega_1}{\sigma_1^2}
+\frac{E_n-\omega_2}{\sigma_2^2}-i\tau
\right)/\sigma_t^2.
\end{eqnarray}
The residual integral in~\eqref{eq:FrequencyIntegral} can be expressed in closed form in terms of the Faddeyeva function of complex argument, $\mathcal{W}(z)$, which, in the upper complex semi-plane, admits the following integral representation (see ~\cite[\href{https://dlmf.nist.gov/7.7.E2}{(7.7.2)}]{NIST:DLMF}),
\begin{equation}
\mathcal{W}(z) \equiv \frac{i}{\pi}\int_{-\infty}^\infty \frac{e^{-t^2}\,dt}{z-t},\qquad \Im z > 0.
\end{equation}
The Faddeyeva function can be extended analytically to the remainder of the complex plane, $\mathcal{W}(z) = 2/e^{z^2}-\mathcal{W}(-z)$. To compute the frequency integral, let us change the integration variable from $\omega$ to $y=\sigma_t(\omega-\tilde{\omega})/\sqrt{2}$,
\begin{equation}
\int_{-\infty}^{\infty} d\omega\,\frac{\exp[-\sigma_t^2(\omega-\tilde{\omega})^2/2]}{\omega+i0^+} = 
\int_{-\infty+i\,\Im z}^{\infty+i\,\Im z} \frac{e^{-y^2}\,dy}{z-i0^+-y},\nonumber
\end{equation}
where $z=-\sigma_t\tilde{\omega}/\sqrt{2}$.

If $\Im z>0$, the pole $z-i0^+$ is below the integration path,
\begin{eqnarray}
\int_{-\infty+i\,\Im z}^{\infty+i\,\Im z} &&\frac{e^{-y^2}\,dy}{z-i0^+-y} = 
\int_{-\infty}^{\infty} \frac{e^{-y^2}\,dy}{z-y} - 2 \,e^{-z^2} \nonumber\\
&=& -i\pi\mathcal{W}(z)+2i\pi\,e^{-z^2}
= i\pi\mathcal{W}(-z).
\end{eqnarray}
The same result is obtained for $\Im z <0$, and hence
\begin{eqnarray}
I_{21}&=& -i\pi
\frac{A_1A_2}{4\sigma_1\sigma_2}\,
e^{-i(\varphi_1+\varphi_2-E_nt_2)}\,\times\\
&\times&e^{-\omega_1^2/2\sigma_1^2-(E_n-\omega_2)^2/2\sigma_2^2
+ \sigma_t^2(\tilde{\omega}-E_p)^2/2}\,\,
\mathcal{W}\left(\frac{\sigma_t\tilde{\omega}}{\sqrt{2}}\right).\nonumber
\end{eqnarray}

%

\end{document}